%% file: ExactChannelSynthesis.tex
\documentclass[10pt]{IEEEtran}
\usepackage[LGR,T1]{fontenc}
\usepackage[latin9]{inputenc}
\usepackage{color}
\usepackage{amsmath}
\usepackage{amsthm}
\usepackage{amssymb}
\usepackage{graphicx}

\makeatletter
\theoremstyle{plain}
\newtheorem{thm}{\protect\theoremname}
\theoremstyle{definition}
\newtheorem{defn}[]{\protect\definitionname}
\theoremstyle{plain}
\newtheorem{prop}[]{\protect\propositionname}
\theoremstyle{remark}
\newtheorem{rem}[]{\protect\remarkname}
\theoremstyle{definition}
\newtheorem{example}[]{\protect\examplename}
\theoremstyle{plain}
\newtheorem{cor}[]{\protect\corollaryname}
\theoremstyle{plain}
\newtheorem{lem}[]{\protect\lemmaname}

\usepackage{cite}

\input{preamble}

\allowdisplaybreaks[1]
\makeatother

\providecommand{\corollaryname}{Corollary}
\providecommand{\definitionname}{Definition}
\providecommand{\examplename}{Example}
\providecommand{\lemmaname}{Lemma}
\providecommand{\propositionname}{Proposition}
\providecommand{\remarkname}{Remark}
\providecommand{\theoremname}{Theorem}

\begin{document}
\title{Exact Channel Synthesis }
\author{Lei Yu and Vincent Y. F. Tan, \IEEEmembership{Senior Member,~IEEE}
\thanks{ This work was supported by a Singapore National Research Foundation (NRF) National Cybersecurity R\&D Grant (R-263-000-C74-281 and NRF2015NCR-NCR003-006) and a Singapore Ministry of Education Tier 2 Grant (R-263-000-C83-112). The first author was also supported by a National Natural Science Foundation of China (NSFC) under Grant (61631017). This paper was presented in part at the  2019 IEEE International Symposium on Information Theory (ISIT).} \thanks{ L.~Yu is with the Department of Electrical and Computer Engineering, National University of Singapore (NUS), Singapore 117583 (e-mail: leiyu@nus.edu.sg). V.~Y.~F.~Tan is with the Department of Electrical and Computer Engineering and the Department of Mathematics, NUS, Singapore 119076 (e-mail: vtan@nus.edu.sg).} \thanks{ Communicated by A. Gohari, Associate Editor for Shannon Theory. } \thanks{Copyright (c) 2019 IEEE. Personal use of this material is permitted. However, permission to use this material for any other purposes must be obtained from the IEEE by sending a request to pubs-permissions@ieee.org.} }
\maketitle
\begin{abstract}
We consider the exact channel synthesis problem. This problem concerns
the determination of the minimum amount of information required to
create exact correlation remotely when there is a certain rate of
randomness shared by two terminals. This problem generalizes an existing
approximate version, in which the generated joint distribution is
required to be close to a target distribution under the total variation
(TV) distance measure (instead being exactly equal to the target distribution).
We provide single-letter inner and outer bounds on the admissible
region of the shared randomness rate and the communication rate for
the exact channel synthesis problem. 
These two bounds coincide for doubly symmetric binary sources. We
observe that for such sources, the admissible rate region for exact
channel synthesis is strictly included in that for the TV-approximate
version. We also extend the exact and TV-approximate channel synthesis
problems to sources with countably infinite alphabets and continuous
sources; the latter includes Gaussian sources. As by-products, lemmas
concerning soft-covering under R\'enyi divergence measures are derived.
\end{abstract}


\section{\label{sec:Introduction}Introduction}

\begin{figure*}
\centering \setlength{\unitlength}{0.06cm} { \begin{picture}(140,35)
\put(-5,10){\vector(1,0){30}} \put(-10,13){%
\mbox{%
$X^{n}\sim\pi_{X}^{n}$%
}} \put(25,4){\framebox(30,12){$P_{W_{n}|X^{n}K_{n}}$}} \put(55,10){\vector(1,0){30}}
\put(65,13){%
\mbox{%
$W_{n}$%
}} \put(85,4){\framebox(30,12){$P_{Y^{n}|W_{n}K_{n}}$}} \put(115,10){\vector(1,0){20}}
\put(120,13){%
\mbox{%
$Y^{n}\sim\pi_{Y|X}^{n}(\cdot|X^{n})$%
}} \put(40,27){\vector(0,-1){11}} \put(-5,30){%
\mbox{%
$K_{n}\sim\mathrm{Unif}[1:e^{nR_{0}}]$%
}} \put(100,27){\vector(0,-1){11}} \put(-5,27){\line(1,0){105}}
\end{picture}}

\caption{\label{fig:dcs}The exact channel synthesis problem. We would like
to design the code  $\left(P_{W_{n}|X^{n}K_{n}},P_{Y^{n}|W_{n}K_{n}}\right)$
such that the induced conditional distribution $P_{Y^{n}|X^{n}}$
satisfies $P_{Y^{n}|X^{n}}=\pi_{Y|X}^{n}$.}
\end{figure*}
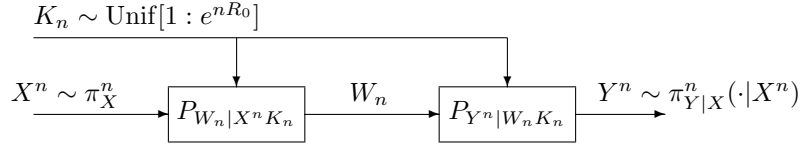

How much information is required to create correlation remotely? This
problem, illustrated in Fig. \ref{fig:dcs} and termed {\em distributed
channel synthesis } (or communication complexity of correlation)
was studied in \cite{bennett2002entanglement,winter2002compression,Cuff,bennett2014quantum,harsha2010communication}.
The exact channel synthesis refers to the problem of determining the
minimum communication rate required to generate a bivariate source
$\left\{ \left(X^{n},Y^{n}\right)\right\} _{n\in\mathbb{N}}$  with
$X^{n}$ generated at the encoder and $Y^{n}$ generated at the decoder
such that the induced joint distribution $P_{X^{n}Y^{n}}$ exactly
equals $\pi_{XY}^{n}$ for all $n\in\mathbb{N}$. In contrast, the
total variation (TV) approximate version of the problem only requires
that the TV distance between $P_{X^{n}Y^{n}}$ and $\pi_{XY}^{n}$
vanishes asymptotically. Bennett et al. \cite{bennett2002entanglement}
studied both exact and TV-approximate syntheses of a target channel.
At almost the same time, Winter \cite{winter2002compression} studied
TV-approximate synthesis of a target channel. However in both these
two works, the authors assumed that unlimited shared randomness is
available at the encoder and decoder. They showed that the minimum
communication rates for both exact and TV-approximate syntheses are
equal to the mutual information $I(X;Y)$ in which $(X,Y)\sim\pi_{XY}$.
Cuff \cite{Cuff} and Bennett et al. \cite{bennett2014quantum} investigated
the tradeoff between the communication rate and the rate of randomness
shared by the encoder and decoder in the TV-approximate simulation
problem. Harsha et al. \cite{harsha2010communication} used a rejection
sampling scheme to study the one-shot version of exact simulation
for discrete $(X,Y)$. They showed that the number of bits of the
shared randomness can be limited to $O(\log\log|\mathcal{X}|+\log|\mathcal{Y}|)$
if the expected description length is increased by $O(\log\left(I(X;Y)+1\right)+\log\log|\mathcal{Y}|)$
bits from the lower bound $I(X;Y)$. Li and El Gamal \cite{li2018strong}
showed that if the expected description length is increased by $\log(I(X;Y)+1)+5$
bits from $I(X;Y)$, then the number of bits of the shared randomness
can be upper bounded by $\log(|\mathcal{X}|(|\mathcal{Y}|-1)+2)$.
Recently, the present authors \cite{yu2018on} considered the exact
channel synthesis problem with no shared randomness and completely
characterized the optimal communication rate for the doubly symmetric
binary source (DSBS). For the DSBS, the present authors observed that
exact channel synthesis requires a strictly larger communication rate
than that required for the TV-approximate version. The tradeoff between
the communication rate and the shared randomness rate for the exact
channel synthesis problem has not been studied, except for the limiting
case of unlimited shared randomness which was studied by Bennett et
al. \cite{bennett2002entanglement}, the limiting case of no shared
randomness which was studied by Kumar, Li, and El Gamal \cite{Kumar}
and the present authors \cite{yu2018on}, as well as the special case
of the symmetric binary erasure source (SBES) which was studied by
Kumar, Li, and El Gamal \cite{Kumar}. In this paper, we study this
problem and make progress on it.

As shown by Bennett et al. \cite{bennett2002entanglement}, when there
exists unlimited shared randomness available at the encoder and decoder,
there exists a scheme to synthesize a target channel if and only if
the asymptotic communication rate is at least the mutual information
$I(X;Y)$. If the sequence of communication rates is restricted to
approach the optimal/minimum rate $I(X;Y)$ asymptotically as $n\to\infty$
(i.e., there is no penalty on the asymptotic communication rate),
then what is the minimum amount of shared randomness required to realize
exact synthesis? Bennett et al. \cite{bennett2014quantum} conjectured
that an exponential number of bits (and hence an infinite rate) of
shared randomness is necessary. For brevity, we term this conjecture
as the BDHSW (Bennett-Devetak-Harrow-Shor-Winter) conjecture. Harsha
et al. \cite{harsha2010communication} (as well as Li and El Gamal
\cite{li2018strong}) disproved this conjecture for $(X,Y)$ with
finite alphabets, and showed that for this case a finite rate (i.e.,
linear number of bits) of shared randomness is sufficient to realize
exact synthesis with no penalty on the asymptotic communication rate.
More precisely, Harsha et al.'s one-shot result implies that the rate
of shared randomness can be upper bounded by $\log|\mathcal{Y}|$.
In this paper, we improve this bound to $H(Y|X)$ and show that our
bound is sharp for the DSBS. We also show that for jointly Gaussian
$(X,Y)$, any finite rate of shared randomness cannot realize exact
synthesis when there is no penalty on the asymptotic communication
rate $I(X;Y)$.

When there is no shared randomness, the channel synthesis problem
reduces to the common information problem. The latter concerns determining
the amount of common randomness required to simulate two correlated
sources in a distributed fashion. The KL-approximate version of such
a problem was first studied by Wyner \cite{Wyner}, who used the normalized
relative entropy (Kullback-Leibler (KL) divergence) to measure the
approximation level (discrepancy) between the simulated joint distribution
and the joint distribution of the original correlated sources. Recently,
the present authors \cite{yu2018wyner,yu2018corrections} generalized
Wyner's result such that the approximation level is measured in terms
of the R\'enyi divergence, thus introducing the notion of R\'enyi common
information. Kumar, Li, and El Gamal \cite{Kumar} considered a variable-length
exact version of Wyner's common information. In their study, in addition
to allowing variable-length codes, they also required the generated
source $(X^{n},Y^{n})\sim\pi_{XY}^{n}$ exactly. For such an exact
synthesis problem, the authors posed an open question as to whether
there exists a bivariate source for which the exact common information
is strictly larger than Wyner's. This question was answered in the
affirmative by the present authors recently \cite{yu2018on}. In \cite{yu2018on},
the present authors completely characterized the exact common information
for the DSBS, and showed that for this source, the exact common information
is strictly larger than Wyner's common information.

Besides the works mentioned above, local TV-approximate simulation
of a channel was studied by Steinberg and Verd\'u \cite{steinberg1994channel};
TV-approximate simulation of a ``bidirectional'' channel via interactive
communication was studied by Yassaee, Gohari, and Aref \cite{yassaee2015channel};
Both the exact and TV-approximate versions of the simulation of a
channel over another noisy channel were studied by Haddadpour, Yassaee,
Beigi, Gohari, and Aref \cite{haddadpour2017simulation}. In particular,
\cite{haddadpour2017simulation} addressed the case of exact simulation
of a binary symmetric channel over a binary erasure channel. The relationship
between the problem of exact channel simulation over another channel
and the problem of zero-error capacity was studied by Cubitt, Leung,
Matthews, and Winter \cite{cubitt2011zero}.

\subsection{Main Contributions}

Our contributions are as follows:
\begin{itemize}
\item First we consider channels with finite input and output alphabets.
We provide a multi-letter characterization on the tradeoff between
or the admissible region of communication rate and shared randomness
rate for exact channel synthesis. Using this multi-letter characterization,
we derive single-letter inner and outer bounds for the admissible
rate region. The inner bound implies that shared randomness with rate
$H(Y|X)$ (or a potentially smaller rate given in \eqref{eq:-31})
suffices to realize exact channel synthesis, even when the sequence
of communication rates is constrained to approach the lowest possible
one $I(X;Y)$ asymptotically. This sharpens Harsha et al.'s upper
bound $\log|\mathcal{Y}|$. 
\item When specialized to the DSBS, the inner and outer bounds coincide.
This implies that the admissible rate region for exact synthesis of
DSBS is completely characterized. Similar to the no shared randomness
case \cite{yu2018on}, when there is shared randomness, the admissible
rate region for exact synthesis is still strictly included in that
for TV-approximate synthesis given by Cuff \cite{Cuff}.
\item We extend the exact and TV-approximate channel synthesis problems
to the synthesis of discrete or continuous channels. In particular,
we provide bounds for jointly Gaussian sources.
\item Concerning proof techniques, we leverage a technique known as mixture
decomposition (or the splitting technique), which was previously used
in \cite{nummelin1978uniform,athreya1978new,roberts2004general,ho2010interplay,Kumar,vellambi2016sufficient}.
However, in this paper (as well as in \cite{yu2018on}), we combine
it with distribution truncation techniques to analyze sources with
countably infinite alphabets. We also combine the mixture decomposition
technique with truncation, discretization, and Li and El Gamal's dyadic
decomposition technique \cite{li2017distributed} to analyze continuous
sources. Furthermore, as by-products of our analyses, various lemmas
that may be of independent interest are derived, e.g., the ``chain
rule for coupling'' lemma, the (distributed and centralized) R\'enyi-covering
lemmas, the log-concavity invariance lemma, etc.
\end{itemize}

\subsection{Notations}

We use $P_{X}$ to denote the probability measure (distribution) of
a random variable $X$ on an alphabet $\mathcal{X}$. For brevity,
we also use $P_{X}$ to denote the corresponding probability mass
function (pmf) for discrete distributions, and the corresponding probability
density function (pdf) for continuous distributions. This will also
be denoted as $P(x)$ (when the random variable $X$ is clear from
the context). We also use $\pi_{X},\widetilde{P}_{X}$, $\widehat{P}_{X}$
and $Q_{X}$ to denote various probability distributions with alphabet
$\mathcal{X}$. The set of probability measures on $\mathcal{X}$
is denoted as $\mathcal{P}\left(\mathcal{X}\right)$, and the set
of conditional probability measures on $\mathcal{Y}$ given a variable
in $\mathcal{X}$ is denoted as $\mathcal{P}(\mathcal{Y}|\mathcal{X}):=\left\{ P_{Y|X}:P_{Y|X}(\cdot|x)\in\mathcal{P}(\mathcal{Y}),x\in\mathcal{X}\right\} $.
Furthermore, the support of a distribution $P\in\calP(\calX)$ is
denoted as $\mathrm{supp}(P)=\{x\in\calX:P(x)>0\}$.

The TV distance between two probability mass functions $P$ and $Q$
with a common alphabet $\calX$ is defined as 
\begin{equation}
|P-Q|:=\frac{1}{2}\sum_{x\in\calX}|P(x)-Q(x)|.
\end{equation}

We use $T_{x^{n}}(x):=\frac{1}{n}\sum_{i=1}^{n}1\left\{ x_{i}=x\right\} $
to denote the type (empirical distribution) of a sequence $x^{n}$,
$T_{X}$ and $V_{Y|X}$ to respectively denote a type of sequences
in $\mathcal{X}^{n}$ and a conditional type of sequences in $\mathcal{Y}^{n}$
(given a sequence $x^{n}\in\calX^{n}$). For a type $T_{X}$, the
type class (set of sequences having the same type $T_{X}$) is denoted
by $\mathcal{T}_{T_{X}}$. For a conditional type $V_{Y|X}$ and a
sequence $x^{n}$, the $V_{Y|X}$-shell of $x^{n}$ (the set of $y^{n}$
sequences having the same conditional type $V_{Y|X}$ given $x^{n}$)
is denoted by $\mathcal{T}_{V_{Y|X}}(x^{n})$. For brevity, sometimes
we use $T(x,y)$ to denote the joint distributions $T(x)V(y|x)$ or
$T(y)V(x|y)$.

The $\epsilon$-strongly and $\epsilon$-weakly typical sets \cite{Gamal,Cover,ho2010information,ho2010markov}
of $P_{X}$ are respectively denoted as 
\begin{align}
\mathcal{T}_{\epsilon}^{\left(n\right)}(P_{X}) & :=\Bigl\{ x^{n}\in\mathcal{X}^{n}:\nonumber \\
 & \quad\left|T_{x^{n}}(x)-P_{X}(x)\right|\leq\epsilon P_{X}(x),\forall x\in\mathcal{X}\Bigr\},\\
\mathcal{A}_{\epsilon}^{(n)}\left(P_{X}\right) & :=\biggl\{ x^{n}\in\mathcal{X}^{n}:\nonumber \\
 & \quad\left|-\frac{1}{n}\log P_{X}^{n}\left(x^{n}\right)-H(X)\right|\leq\epsilon\biggr\}.\label{eq:-22-1}
\end{align}
Note that $\mathcal{T}_{\epsilon}^{\left(n\right)}(P_{X})$ only applies
to sources with finite alphabets. For $\mathcal{A}_{\epsilon}^{(n)}\left(P_{X}\right)$,
if $P_{X}$ is an absolutely continuous distribution, in \eqref{eq:-22-1},
$P_{X}^{n}\left(x^{n}\right)$ and $H(X)$ are respectively replaced
with the corresponding pdf and differential entropy. The corresponding
jointly typical sets are defined similarly. The conditionally $\epsilon$-strongly
typical set of $P_{XY}$ is denoted as 
\begin{equation}
\mathcal{T}_{\epsilon}^{\left(n\right)}(P_{XY}|x^{n}):=\left\{ y^{n}\in\mathcal{Y}^{n}:(x^{n},y^{n})\in\mathcal{T}_{\epsilon}^{\left(n\right)}(P_{XY})\right\} ,
\end{equation}
and the conditionally $\epsilon$-weakly typical set is defined similarly.
For brevity, sometimes we denote $\mathcal{T}_{\epsilon}^{\left(n\right)}(P_{X})$
and $\mathcal{A}_{\epsilon}^{(n)}\left(P_{X}\right)$ as $\mathcal{T}_{\epsilon}^{\left(n\right)}$
and $\mathcal{A}_{\epsilon}^{(n)}$, respectively.

Fix distributions $P_{X},Q_{X}\in\calP(\calX)$. The {\em relative
entropy} and the {\em R\'enyi divergence of order $\infty$} are
respectively defined as 
\begin{align}
D(P_{X}\|Q_{X}) & :=\sum_{x\in\mathrm{supp}(P_{X})}P_{X}(x)\log\frac{P_{X}(x)}{Q_{X}(x)}\label{eq:-19-2}\\*
D_{\infty}(P_{X}\|Q_{X}) & :=\log\sup_{x\in\mathrm{supp}(P_{X})}\frac{P_{X}(x)}{Q_{X}(x)},\label{eq:-40}
\end{align}
and the conditional versions are respectively defined as 
\begin{align}
D(P_{Y|X}\|Q_{Y|X}|P_{X}) & :=D(P_{X}P_{Y|X}\|P_{X}Q_{Y|X})\\*
D_{\infty}(P_{Y|X}\|Q_{Y|X}|P_{X}) & :=D_{\infty}(P_{X}P_{Y|X}\|P_{X}Q_{Y|X}),
\end{align}
where the summations in \eqref{eq:-19-2} and \eqref{eq:-40} are
taken over the elements in $\mathrm{supp}(P_{X})$. Throughout, $\log$
and $\exp$ are to the natural base $e$ .

Denote the coupling set of $(P_{X},P_{Y})$ as 
\begin{align}
C(P_{X},P_{Y}) & :=\left\{ Q_{XY}\in\mathcal{P}(\mathcal{X}\times\mathcal{Y}):Q_{X}=P_{X},Q_{Y}=P_{Y}\right\} .
\end{align}
For $i,j\in\mathbb{Z}$, and $i\le j$, we define $[i:j]:=\{i,i+1,\ldots,j\}$.
Given a number $a\in[0,1]$, we define $\overline{a}=1-a$. For any
real number $c$ and any set $\mathcal{A}\subseteq\mathbb{R}^{n}$,
we define $c\mathcal{A}:=\left\{ ca:\,a\in\mathcal{A}\right\} $.
For a set $\mathcal{A}\subseteq\mathbb{R}^{n}$, we use $\mathrm{cl}\,\mathcal{A}$
and $\mathrm{int}\,\mathcal{A}$ to denote the closure and interior
of $\mathcal{A}$ respectively. For a sequence $\left\{ \mathcal{A}_{n}\right\} $
of subsets of a space, $\limsup_{n\to\infty}\mathcal{A}_{n}:=\bigcap_{n\geq1}\bigcup_{j\geq n}\mathcal{A}_{j}$.

We say that a sequence of real numbers $\left(a_{n}\right)$ converges
to a finite real value $a$ (at least) exponentially fast if there
exist a real number $b>1$ and a positive integer $N$ such that $\left|a_{n}-a\right|\le b^{-n}$
for all $n\ge N$. We say that a sequence of real numbers $\left(a_{n}\right)$
converges to a finite real value $a$ (at least) doubly exponentially
fast if there exist real numbers $b,c>1$ and a positive integer $N$
such that $\left|a_{n}-a\right|\le b^{-c^{n}}$ for all $n\ge N$.

For two distributions $P$ and $Q$ defined on the same measurable
space, we use $P\ll Q$ to denote that $P$ is absolutely continuous
with respect to $Q$. If $P\ll Q$, we use $\frac{\mathrm{d}P}{\mathrm{d}Q}$
to denote the Radon--Nikodym derivative of $P$ with respect to $Q$.

\section{Problem Formulation}

Consider the distributed source simulation setup depicted in Fig.
\ref{fig:dcs}. A sender and a receiver share a uniformly distributed
source of randomness\footnote{For simplicity, we assume that $e^{nR}$ and similar expressions are
integers.} $K_{n}\sim\mathrm{Unif}\left(\calK_{n}\right),\calK_{n}:=[1:e^{nR_{0}}]$.
The sender has access to a memoryless source $X^{n}\sim\pi_{X}^{n}$
that is independent of $K_{n}$, and wants to transmit information
about the correlation between correlated sources $\left(X^{n},Y^{n}\right)\sim\pi_{XY}^{n}$
to the receiver. Here we assume $\supp(\pi_{XY})\subseteq\mathcal{X}\times\mathcal{Y}$
but not necessarily $\supp(\pi_{XY})=\mathcal{X}\times\mathcal{Y}$.
Given the shared randomness and the correlation information from the
sender, the receiver generates a memoryless source $Y^{n}\sim\pi_{Y|X}^{n}(\cdot|X^{n})$.
Specifically, given $X^{n}$ and $K_{n}$, the sender generates a
``message'' (i.e., a discrete random variable) $W_{n}$ by a random
mapping $P_{W_{n}|X^{n}K_{n}}$, and then sends it to the receiver
error free. Upon accessing to $K_{n}$ and receiving $W_{n}$, the
receiver generates a source $Y^{n}$ by a random mapping $P_{Y^{n}|W_{n}K_{n}}$.
Now we would like to determine the minimum amount of communication
such that the joint distribution of $\left(X^{n},Y^{n}\right)$ is
$\pi_{XY}^{n}$. Next we provide a precise formulation of this problem.

Define $\left\{ 0,1\right\} ^{*}:=\bigcup_{n\ge1}\left\{ 0,1\right\} ^{n}$
as the set of finite-length strings of symbols from a binary alphabet
$\left\{ 0,1\right\} $. Denote the alphabet of the random variable
$W_{n}$ as ${\cal W}_{n}$, which is a countable set. Consider a
set of (more precisely, a sequence of) prefix-free codes \cite{Cover},
$\boldsymbol{f}=\left\{ f_{k}:k\in{\cal K}_{n}\right\} $, which consists
of $f_{k}:{\cal W}_{n}\to\left\{ 0,1\right\} ^{*},k\in{\cal K}_{n}$.
Then for each pair $\left(w,k\right)\in{\cal W}_{n}\times{\cal K}_{n}$
and the set of codes $\boldsymbol{f}$, let $\ell_{\boldsymbol{f}}(w|k)$
denote the length of the codeword $f_{k}\left(w\right)$, where $f_{k}$
is the $k$-th component of $\boldsymbol{f}$.
\begin{defn}
The expected codeword length $L_{\boldsymbol{f}}$ for compressing
the random variable $W_{n}$ given $K_{n}$ by a prefix-free code
set $\boldsymbol{f}$ is denoted as $L_{\boldsymbol{f}}(W_{n}|K_{n}):=\mathbb{E}\left[\ell_{\boldsymbol{f}}(W_{n}|K_{n})\right]$.
Here the expectation is taken respect to the random variables $\left(W_{n},K_{n}\right)$.
\end{defn}
\begin{defn}
A variable-length  $(n,R_{0},R)$-code consists of a pair of random
mappings $P_{W_{n}|X^{n}K_{n}}:\mathcal{X}^{n}\times\calK_{n}\to{\cal W}_{n},P_{Y^{n}|W_{n}K_{n}}:{\cal W}_{n}\times\calK_{n}\to\mathcal{Y}^{n}$
and a set of prefix-free codes $\boldsymbol{f}=\left\{ f_{k}:{\cal W}_{n}\to\left\{ 0,1\right\} ^{*}\right\} _{k\in{\cal K}_{n}}$
for some countable set ${\cal W}_{n}$ such that\footnote{Our results in this paper still hold if we replace ${L_{\boldsymbol{f}}(W_{n}|K_{n})}/{n}\le R$
with a stronger constraint ${L_{\boldsymbol{f}}(W_{n}|k)}/{n}\le R$
for all $k$, where ${L_{\boldsymbol{f}}(W_{n}|k)}:=\mathbb{E}_{W_{n}|K_{n}=k}\left[\ell_{\boldsymbol{f}}(W_{n}|k)\right]$.} the expected codeword length for $\left(W_{n},K_{n}\right)$ satisfies
${L_{\boldsymbol{f}}(W_{n}|K_{n})}/{n}\le R$, where $\left(W_{n},K_{n}\right)$
is distributed as 
\begin{equation}
P_{W_{n}K_{n}}(w,k)=\sum_{x^{n}\in\mathcal{X}^{n}}\frac{1}{|{\cal K}_{n}|}\pi_{X}^{n}\left(x^{n}\right)P_{W_{n}|X^{n}K_{n}}(w|x^{n},k).
\end{equation}
\end{defn}
By using such synthesis codes, $W_{n}$ can be transmitted from the
sender to the receiver without error. Hence the generated (or synthesized)
distribution for such setting is 
\begin{align}
 & P_{Y^{n}|X^{n}}(y^{n}|x^{n}):=\sum_{\left(w,k\right)\in{\cal W}_{n}\times{\cal K}_{n}}\frac{1}{|{\cal K}_{n}|}\nonumber \\
 & \qquad\times P_{W_{n}|X^{n}K_{n}}(w|x^{n},k)P_{Y^{n}|W_{n}K_{n}}(y^{n}|w,k),\label{eq:-74}
\end{align}
which is required to be equal to $\pi_{Y|X}^{n}$ exactly. It is worth
noting that under the assumption $K_{n}\sim\mathrm{Unif}\left(\calK_{n}\right)$,
the synthesized channel $P_{Y^{n}|X^{n}}$ is determined only by the
random mapping pair $\left(P_{W_{n}|X^{n}K_{n}},P_{Y^{n}|W_{n}K_{n}}\right)$,
and does not depend on the source distribution $\pi_{X}^{n}$. However,
the code rate induced by a given synthesis code indeed depends on
$\pi_{X}^{n}$. Given the shared randomness rate $R_{0}$, the minimum
asymptotic communication rate required to ensure $P_{Y^{n}|X^{n}}=\pi_{Y|X}^{n}$
for all $n\ge1$ is $\limsup_{n\to\infty}\frac{1}{n}{L_{\boldsymbol{f}}(W_{n}|K_{n})}$. 
\begin{defn}
The admissible region of shared randomness rate and communication
rate for the exact channel synthesis problem is defined as 
\begin{align}
 & \mathcal{R}_{\mathrm{Exact}}(\pi_{XY})\nonumber \\
 & :=\mathrm{cl}\left\{ \begin{array}{l}
(R_{0},R)\in\mathbb{R}_{\ge0}^{2}:\\
\exists\left\{ \textrm{variable-length }(n,R_{0},R^{(n)})\textrm{ code}\right\} _{n=1}^{\infty}\textrm{ s.t.}\\
\qquad P_{Y^{n}|X^{n}}=\pi_{Y|X}^{n},\forall n,\\
\qquad R\ge\limsup_{n\to\infty}R^{(n)}
\end{array}\right\} .
\end{align}
\end{defn}
Observe that $L_{\boldsymbol{f}}(W_{n}|K_{n})=\mathbb{E}_{K_{n}}\mathbb{E}_{W_{n}|K_{n}}\left[\ell_{\boldsymbol{f}}(W_{n}|K_{n})\right]$.
Hence to minimize the  expected codeword length $L_{\boldsymbol{f}}(W_{n}|K_{n})$,
it suffices to minimize $\mathbb{E}\left[\ell_{\boldsymbol{f}}(W_{n}|K_{n})|K_{n}=k\right]$
for each $k$. For each $k\in{\cal K}_{n}$, we use an optimal prefix-free
code (e.g., a Huffman code) $f_{k}$ to compress $W_{n}$. The resulting
expected codeword length given $K_{n}=k$ satisfies $H(W_{n}|K_{n}=k)\leq\mathbb{E}_{W_{n}|K_{n}=k}\left[\ell_{\boldsymbol{f}}(W_{n}|k)\right]<H(W_{n}|K_{n}=k)+1$
\cite[Theorem 5.4.1]{Cover}. Hence for a set of optimal prefix-free
codes $\boldsymbol{f}^{*}=\left\{ f_{k}:k\in{\cal K}_{n}\right\} $,
the expected codeword length also satisfies 
\begin{equation}
H(W_{n}|K_{n})\leq L_{\boldsymbol{f}^{*}}(W_{n}|K_{n})<H(W_{n}|K_{n})+1.\label{eq:-16}
\end{equation}
Consequently, 
\begin{equation}
\frac{1}{n}L_{\boldsymbol{f}^{*}}(W_{n}|K_{n})-\frac{1}{n}H(W_{n}|K_{n})\to0\textrm{ as }n\to\infty.\label{eq:-17}
\end{equation}
Based on such an argument, we provide the following multi-letter characterization
for $\mathcal{R}_{\mathrm{Exact}}(\pi_{XY})$ as follows. The proof
is given in Appendix \ref{sec:entropycharacterization}. 
\begin{prop}
\label{prop:entropycharacterization} We have
\begin{align}
 & \mathcal{R}_{\mathrm{Exact}}(\pi_{XY})\nonumber \\
 & =\mathrm{cl}\bigcup_{n\ge1}\left\{ \begin{array}{l}
(R_{0},R):\exists\left(P_{W_{n}|X^{n}K_{n}},P_{Y^{n}|W_{n}K_{n}}\right)\textrm{ s.t.}\\
\qquad P_{Y^{n}|X^{n}}=\pi_{Y|X}^{n},\\
\qquad R\ge\frac{1}{n}H(W_{n}|K_{n})
\end{array}\right\} .\label{eq:}
\end{align}
\end{prop}
This multi-letter characterization does not depend on the set of prefix-free
codes $\boldsymbol{f}$. Hence a variable-length synthesis code can
be represented by a pair of random mappings $(P_{W_{n}|X^{n}K_{n}},P_{Y^{n}|W_{n}K_{n}})$,
where the dependence on $\boldsymbol{f}$ is omitted.

\section{Main Results for Distributions with Finite Alphabets}

In this section, we assume that $\pi_{XY}$ has a finite alphabet.
We first introduce a new quantity, the maximal cross-entropy, use
it to provide a multi-letter expression for the exact channel synthesis
problem. Based on such an expression, we then derive single-letter
inner and outer bounds. Finally, we solve the exact synthesis problem
for the DSBS.

\subsection{\label{subsec:Maximal-Cross-Entropy}Maximal Cross-Entropy}
\begin{defn}
For a distribution tuple $\left(P_{X},P_{Y},\pi_{XY}\right)$, define
the maximal cross-entropy over couplings in $C(P_{X},P_{Y})$ as 
\begin{align}
 & \mathcal{H}(P_{X},P_{Y}\|\pi_{XY})\nonumber \\
 & :=\max_{P_{XY}\in C(P_{X},P_{Y})}\sum_{x,y}P_{XY}(x,y)\log\frac{1}{\pi\left(x,y\right)}.\label{eq:maximalcrossentropy}
\end{align}
\end{defn}
\begin{rem}
The concept of maximal (relative) cross-entropy can be easily generalized
to distributions with arbitrary alphabets by rewriting the RHS of
\eqref{eq:maximalcrossentropy} as $\sup_{P_{XY}\in C(P_{X},P_{Y})}-\mathbb{E}_{P_{XY}}\log\frac{\mathrm{d}\pi}{\mathrm{d}\mu}\left(X,Y\right)$,
where $\mu$ denotes a reference measure such that $\pi\ll\mu$. 
\end{rem}
The coupling set and the maximal cross-entropy have the following
intuitive interpretations.  Assume the alphabets $\mathcal{X}$ and
$\mathcal{Y}$ are finite. Consider a joint distribution $\pi_{XY}$,
a pair of distributions $\left(P_{X},P_{Y}\right)$, and a sequence
of pairs of types 
\begin{equation}
\left\{ \left(T_{X}^{(n)},T_{Y}^{(n)}\right)\in\mathcal{P}_{n}\left(\mathcal{X}\right)\times\mathcal{P}_{n}\left(\mathcal{Y}\right)\right\} _{n\in\mathbb{N}}
\end{equation}
such that $\left(T_{X}^{(n)},T_{Y}^{(n)}\right)\to\left(P_{X},P_{Y}\right)$
as $n\to\infty$. Then the sets of the joint types of $\left(x^{n},y^{n}\right)$
such that $T_{x^{n}}=T_{X}^{(n)},T_{y^{n}}=T_{Y}^{(n)}$ satisfy 
\begin{align}
 & \mathrm{cl\,}\limsup_{n\to\infty}\left\{ T_{x^{n},y^{n}}:\exists\left(x^{n},y^{n}\right)\textrm{ s.t. }T_{x^{n}}=T_{X}^{(n)},T_{y^{n}}=T_{Y}^{(n)}\right\} \nonumber \\
 & \qquad=C(P_{X},P_{Y}).
\end{align}
The exponents of probabilities $\pi_{XY}^{n}\left(x^{n},y^{n}\right)$
such that $T_{x^{n}}=T_{X}^{(n)},T_{y^{n}}=T_{Y}^{(n)}$ satisfy that
\begin{align}
 & \lim_{n\to\infty}\min_{\substack{\left(x^{n},y^{n}\right):\\
T_{x^{n}}=T_{X}^{(n)},\\
T_{y^{n}}=T_{Y}^{(n)}
}
}-\frac{1}{n}\log\pi_{XY}^{n}\left(x^{n},y^{n}\right)\nonumber \\
 & =\lim_{n\to\infty}\min_{\substack{\left(x^{n},y^{n}\right):\\
T_{x^{n}}=T_{X}^{(n)},\\
T_{y^{n}}=T_{Y}^{(n)}
}
}\sum_{x,y}T_{x^{n},y^{n}}(x,y)\log\frac{1}{\pi\left(x,y\right)}\\
 & =\mathcal{H}(P_{X},P_{Y}\|\pi_{XY}).
\end{align}

Furthermore, the following fundamental properties on maximal cross-entropy
hold. The proof  is provided in Appendix \ref{sec:crossentropy}. 
\begin{prop}
\label{prop:maximalcrossentropy}Assume the alphabets $\mathcal{X}$
and $\mathcal{Y}$ are finite. Let $\pi_{XY}$ be a joint distribution
with marginals $\pi_{X}$ and $\pi_{Y}$. a) Then we have 
\begin{equation}
\mathcal{H}(\pi_{X},\pi_{Y}\|\pi_{XY})\geq H(\pi_{XY}),\label{eq:maximalcrossentropy-1}
\end{equation}
where equality in \eqref{eq:maximalcrossentropy-1} holds if and only
if $\pi_{XY}=\pi_{X}\pi_{Y}$. b) Moreover, assume $\supp(\pi_{XY})=\mathcal{X}\times\mathcal{Y}$.
Then for any distributions $P_{X}$ and $P_{Y}$ such that $\supp(P_{X})=\mathcal{X},\supp(P_{Y})=\mathcal{Y}$,
we have 
\begin{equation}
\mathcal{H}(P_{X},P_{Y}\|\pi_{XY})\geq\sum_{x,y}P_{X}(x)P_{Y}(y)\log\frac{1}{\pi\left(x,y\right)},\label{eq:maximalcrossentropy-1-1-1}
\end{equation}
where equality in \eqref{eq:maximalcrossentropy-1-1-1} holds if and
only if $\pi_{XY}=\pi_{X}\pi_{Y}$.
\end{prop}

\begin{example}[DSBS]
\label{exa:-Consider-a} Consider a DSBS $\left(X,Y\right)$ with
distribution 
\begin{equation}
\pi_{XY}=\left[\begin{array}{cc}
\alpha_{0} & \beta_{0}\\
\beta_{0} & \alpha_{0}
\end{array}\right]\label{eq:-10-1}
\end{equation}
where $\alpha_{0}=\frac{1-p}{2},\beta_{0}=\frac{p}{2}$ with $p\in[0,\frac{1}{2}]$.
Here w.l.o.g., we restrict $p\in[0,\frac{1}{2}]$, since otherwise,
we can set $X\oplus1$ to $X$ and the same conclusions follow. Consider
\begin{equation}
P_{X}=\left(\alpha,\overline{\alpha}\right),\quad P_{Y}=\left(\beta,\overline{\beta}\right)
\end{equation}
for some $\alpha,\beta\in[0,1]$. Then 
\begin{align}
 & \mathcal{H}(P_{X},P_{Y}\|\pi_{XY})\nonumber \\
 & =\log\frac{1}{\alpha_{0}}+\left(\min\{\alpha,\overline{\beta}\}+\min\{\overline{\alpha},\beta\right)\log\frac{\alpha_{0}}{\beta_{0}}\\
 & =\log\frac{1}{\alpha_{0}}+\min\{\alpha+\beta,\overline{\alpha}+\overline{\beta}\}\log\frac{\alpha_{0}}{\beta_{0}}.
\end{align}
Furthermore, if $P_{X}=\pi_{X},P_{Y}=\pi_{Y}$ (i.e., $\alpha=\beta=\frac{1}{2}$
), then 
\begin{align}
\mathcal{H}(\pi_{X},\pi_{Y}\|\pi_{XY}) & =\log\frac{1}{\beta_{0}}.
\end{align}
In contrast, 
\begin{align}
H(\pi_{XY}) & =2\alpha_{0}\log\frac{1}{\alpha_{0}}+2\beta_{0}\log\frac{1}{\beta_{0}}\\
 & \leq\mathcal{H}(\pi_{X},\pi_{Y}\|\pi_{XY}),\label{eq:-44}
\end{align}
where equality in \eqref{eq:-44} holds if and only if $p=\frac{1}{2}$. 
\end{example}
\begin{example}[Gaussian Source]
\label{exa:-Consider-a-1} Consider a bivariate Gaussian source $\pi_{XY}=\mathcal{N}\left(0,\Sigma_{XY}\right)$
where $\Sigma_{XY}=\left[\begin{array}{cc}
1 & \rho\\
\rho & 1
\end{array}\right]$ with correlation coefficient $\rho\in[0,1)$. Here without loss of
any generality, we assume the correlation coefficient $\rho$ between
$(X,Y)$ is nonnegative; otherwise, we can set $-X$ to $X$. Consider
\begin{equation}
P_{X}=\mathcal{N}(\mu_{1},\alpha),\quad P_{Y}=\mathcal{N}(\mu_{2},\beta)
\end{equation}
for some $\alpha,\beta>0$. Then\footnote{Here computing $\min_{P_{XY}\in C(P_{X},P_{Y})}\mathbb{E}\left[XY\right]$
is equivalent to computing the Wasserstein distance of order 2 $\mathsf{W}_{2}(P_{X},P_{Y'}):=\min_{P_{XY'}\in C(P_{X},P_{Y'})}\mathbb{E}[\left(X-Y'\right)^{2}]$
where $Y':=-Y$. It is well known \cite[Example 3.2.14]{rachev1998mass}
that for $P_{X},P_{Y'}$ defined on $\mathbb{R}$, $\mathsf{W}_{2}(P_{X},P_{Y'})=\mathbb{E}[(F_{X}^{-1}(U)-F_{Y}^{-1}(U))^{2}]$,
where $U$ is a uniform random variable on $\left[0,1\right]$, and
$F_{X}^{-1}(u):=\inf\{x\in\mathbb{R}:F_{X}(x)\geq u\}$  denotes
the generalized inverses of the cumulative distribution functions
 $F_{X}$  of $P_{X}$ (and $F_{Y}^{-1}(\cdot)$ is defined similarly).} 
\begin{align}
 & \mathcal{H}(P_{X},P_{Y}\|\pi_{XY})\nonumber \\
 & =\log\left(2\pi\sqrt{1-\rho^{2}}\right)+\frac{1-\rho\min_{P_{XY}\in C(P_{X},P_{Y})}\mathbb{E}\left[XY\right]}{1-\rho^{2}}\\
 & =\log\left(2\pi\sqrt{1-\rho^{2}}\right)+\frac{1}{1-\rho^{2}}-\frac{\rho}{1-\rho^{2}}\nonumber \\
 & \qquad\times\left(\min_{P_{XY}\in C(P_{X},P_{Y})}\mathbb{E}\left[\left(X-\mu_{1}\right)\left(Y-\mu_{2}\right)\right]+\mu_{1}\mu_{2}\right)\\
 & =\log\left(2\pi\sqrt{1-\rho^{2}}\right)+\frac{1+\rho\left(\sqrt{\alpha\beta}-\mu_{1}\mu_{2}\right)}{1-\rho^{2}},\label{eq:-45}
\end{align}
where \eqref{eq:-45} follows by the Cauchy--Schwarz inequality.
Furthermore, if $P_{X}=\pi_{X},P_{Y}=\pi_{Y}$ (i.e., $\mu_{1}=\mu_{2}=0,\alpha=\beta=1$
), then 
\begin{align}
\mathcal{H}(\pi_{X},\pi_{Y}\|\pi_{XY}) & =\log\left(2\pi\sqrt{1-\rho^{2}}\right)+\frac{1}{1-\rho}.
\end{align}
In contrast, 
\begin{equation}
H(\pi_{XY})=\log\left(2\pi e\sqrt{1-\rho^{2}}\right)\leq\mathcal{H}(\pi_{X},\pi_{Y}\|\pi_{XY}),
\end{equation}
where equality holds if and only if $\rho=0$.
\end{example}

\subsection{Multi-Letter Characterization}

Based on the maximal cross-entropy defined above, we characterize
the admissible rate region $\mathcal{R}_{\mathrm{Exact}}(\pi_{XY})$
by using multi-letter expressions. The proof of Theorem \ref{thm:multiletter}
is given in Appendix \ref{sec:equivalence}. 
\begin{thm}[Multi-letter Characterization]
\label{thm:multiletter} For a joint distribution $\pi_{XY}$ defined
on a finite alphabet, 
\begin{equation}
\mathcal{R}_{\mathrm{Exact}}(\pi_{XY})=\mathrm{cl}\,\bigcup_{n\ge1}\frac{1}{n}\mathcal{R}(\pi_{XY}^{n}),
\end{equation}
where 
\begin{align}
 & \mathcal{R}(\pi_{XY}^{n}):=\nonumber \\
 & \left\{ \begin{array}{rcl}
\left(R_{0},R\right) & : & \exists P_{W}P_{X|W}P_{Y|W}\textrm{ s.t. }\\
P_{XY} & = & \pi_{XY},\\
R & \ge & I(W;X^{n}),\\
R_{0}+R & \ge & -H(X^{n}Y^{n}|W)+\sum_{w}P(w)\\
 &  & \times\mathcal{H}(P_{X^{n}|W=w},P_{Y^{n}|W=w}\|\pi_{XY}^{n})
\end{array}\right\} .\label{eq:-30}
\end{align}
\end{thm}
In our achievability scheme, we apply a mixture decomposition technique
(or known as the splitting technique) to construct a variable-length
exact synthesis code. This code can be thought of as a mixture of
a fixed-length $\infty$-R\'enyi-approximate code and a completely lossless
code. The $\infty$-R\'enyi-approximate code is a fixed-length code
which generates a channel $P_{Y^{n}|X^{n}}$ that approaches the target
channel $\pi_{Y|X}^{n}$ asymptotically under the $\infty$-R\'enyi
divergence measure. Based on the channel $P_{Y^{n}|X^{n}}$, we can
decompose $\pi_{Y|X}^{n}$ as a mixture conditional distribution 
\begin{equation}
\pi_{Y|X}^{n}=e^{-\delta_{n}}P_{Y^{n}|X^{n}}+\left(1-e^{-\delta_{n}}\right)\widehat{P}_{Y^{n}|X^{n}},
\end{equation}
for some asymptotically vanishing sequence $\left(\delta_{n}\right)$
where 
\begin{equation}
\widehat{P}_{Y^{n}|X^{n}}:=\frac{e^{\delta_{n}}\pi_{Y|X}^{n}-P_{Y^{n}|X^{n}}}{e^{\delta_{n}}-1}.
\end{equation}
For the ``residual channel'' $\widehat{P}_{Y^{n}|X^{n}}$, we adopt
a completely lossless code to synthesize it. In this code, upon observing
$x^{n}$, the sender generates a random sequence $Y^{n}\sim\widehat{P}_{Y^{n}|X^{n}=x^{n}}$,
and then compresses $Y^{n}$ by using a prefix-free code with rate
$\leq\log|\mathcal{Y}|$. In our scheme, the lossless code is invoked
with asymmetrically vanishing probability $1-e^{-\delta_{n}}$, and
hence the performance of our scheme is dominated by the $\infty$-R\'enyi-approximate
code which requires a much lower rate. The $\infty$-R\'enyi-approximate
code we adopt is a truncated i.i.d.\ code. For such a code, the codewords
are independent and each codeword is drawn according to a distribution
$P_{W^{n}}$ which is generated by truncating a product distribution
$Q_{W}^{n}$ onto some (strongly) typical set. Truncated i.i.d. codes
are rather useful for $\infty$-R\'enyi-approximate synthesis (but not
for TV-approximate synthesis). This follows from the following argument.
Observe that for both $\infty$-R\'enyi-approximate synthesis and TV-approximate
synthesis, $X^{n}\to W_{n}K_{n}\to Y^{n}$ forms a Markov chain. Hence
given $\left(W_{n},K_{n}\right)=\left(w,k\right)$, the support of
$P_{X^{n}|W_{n}K_{n}}\left(\cdot|w,k\right)P_{Y^{n}|W_{n}K_{n}}\left(\cdot|w,k\right)$
is a product set, which in turn implies that the support of $P_{X^{n}Y^{n}}$
is the union of a family of product sets. Such a requirement leads
to the fact that the support of $P_{X^{n}Y^{n}}$ includes not only
a jointly typical set, but also other joint type classes (which is
termed by us as the type overflow phenomenon). TV-approximate synthesis
only requires the sequences in a typical set to be well-simulated.
However, $\infty$-R\'enyi-approximate synthesis requires all the sequences
in the support of $P_{X^{n}Y^{n}}$ to be well-simulated. Hence type
overflow does not affect TV-approximate synthesis, but plays a crucial
role for $\infty$-R\'enyi-approximate synthesis (or exact synthesis).
Truncated i.i.d. coding is an efficient approach to control the possible
types of the output sequence of a code (or more precisely, to mitigate
the effects of type overflow). Furthermore, truncated i.i.d. codes
have also been used by the present authors \cite{yu2019renyi,yu2018wyner,yu2018corrections,yu2018on}
to study the R\'enyi and exact common informations, and by Vellambi
and Kliewer \cite{vellambi2016sufficient,vellambi2018new} to study
sufficient conditions for the equality of the exact and Wyner's common
informations. 

Based on the type overflow argument given above and the intuitive
explanation of the maximal cross-entropy given in Subsection \ref{subsec:Maximal-Cross-Entropy},
our bounds are easy to comprehend intuitively. For simplicity, we
only consider the single-letter expression $\mathcal{R}(\pi_{XY})$.
The first inequality follows from the fact that lossless transmission
of the message $W_{n}$ requires rate at least $I(W;X)$. The second
inequality follows from the following argument. The exact channel
synthesis requires that there exists a sequence of variable-length
codes with rates $\left(R_{0},R\right)$ satisfying $\frac{P_{Y^{n}|X^{n}}(y^{n}|x^{n})}{\pi_{Y|X}^{n}(y^{n}|x^{n})}=1$
for all $(x^{n},y^{n})\in\mathcal{X}^{n}\times\mathcal{Y}^{n}$. By
using the mixture decomposition technique, the exact channel synthesis
problem can be relaxed to the $\infty$-R\'enyi-approximate synthesis
problem, which requires that there exists a sequence of fixed-length
codes with rates $\left(R_{0},R\right)$ satisfying 
\begin{equation}
\frac{\pi_{X}^{n}(x^{n})}{P_{X^{n}}(x^{n})}=1+o(1)\label{eq:-33}
\end{equation}
for all $x^{n}\in\mathcal{T}_{\epsilon}^{\left(n\right)}(\pi_{X})$
and 
\begin{equation}
\frac{P_{X^{n}Y^{n}}(x^{n},y^{n})}{\pi_{XY}^{n}(x^{n},y^{n})}\leq1+o(1)\label{eq:-35}
\end{equation}
for all $(x^{n},y^{n})\in\supp\left(P_{X^{n}Y^{n}}\right)$; see Lemma
\ref{lem:sufficiency}. The requirement \eqref{eq:-33} is satisfied,
as long as $R\geq I(W;X)$. As for the constraint in \eqref{eq:-35},
observe that by using truncated i.i.d. codes, to mitigate the effect
of type overflow we can restrict $(W^{n},X^{n})\in\mathcal{T}_{\epsilon}^{\left(n\right)}\left(P_{WX}\right)$
and $(W^{n},Y^{n})\in\mathcal{T}_{\epsilon}^{\left(n\right)}\left(P_{WY}\right)$.
Assume that $M_{n}$ is the message for $\infty$-R\'enyi-approximate
synthesis. Roughly speaking, for a given $R_{0}$, a sequence of optimal
codes that achieves the minimum asymptotically communication rate
satisfies the following ``property'': Each pair of output sequences
$(x^{n},y^{n})$ is only covered by less than $e^{n\delta}$ codewords
for any $\delta>0$ (otherwise, the code rate $R$ can be further
reduced). This ``property'' implies that for sufficiently large
$n$ and sufficiently small $\epsilon$, 
\begin{align}
 & P_{X^{n}Y^{n}}(x^{n},y^{n})\nonumber \\
 & \approx P_{M_{n}K_{n}}(m,k)P_{X|W}^{n}(x^{n}|w^{n}(m,k))P_{Y|W}^{n}(y^{n}|w^{n}(m,k))\\
 & \approx e^{-n\left(R+R_{0}\right)}e^{-nH(X|W)}e^{-nH(Y|W)}.\label{eq:-38}
\end{align}
On the other hand,
\begin{align}
 & \min_{(x^{n},y^{n})\in\supp\left(P_{X^{n}Y^{n}}\right)}\pi_{XY}^{n}(x^{n},y^{n})\nonumber \\
 & \approx\min_{\left(w^{n},x^{n},y^{n}\right):T_{w^{n}x^{n}}\approx P_{WX},T_{w^{n}y^{n}}\approx P_{WY}}\pi_{XY}^{n}(x^{n},y^{n})\\
 & \approx e^{-n\sum_{w}P_{W}(w)\mathcal{H}(P_{X|W=w},P_{Y|W=w}\|\pi_{XY})}.\label{eq:-39}
\end{align}
Substituting \eqref{eq:-38} and \eqref{eq:-39} into \eqref{eq:-35},
we obtain 
\begin{align}
 & R_{0}+R\nonumber \\
 & \gtrsim-H(XY|W)+\sum_{w}P(w)\mathcal{H}(P_{X|W=w},P_{Y|W=w}\|\pi_{XY}).
\end{align}
This is the second constraint in \eqref{eq:-30}.

\subsection{Single-letter Bounds}

Define 
\begin{align}
 & \mathcal{R}_{\mathrm{Cuff}}(\pi_{XY})\nonumber \\
 & :=\left\{ \begin{array}{rcl}
\left(R_{0},R\right) & : & \exists P_{W}P_{X|W}P_{Y|W}\textrm{ s.t. }\\
P_{XY} & = & \pi_{XY},\\
R & \ge & I(W;X),\\
R_{0}+R & \ge & I\left(W;XY\right)
\end{array}\right\} ,\label{eq:-TV}
\end{align}
\begin{align}
 & \mathcal{R}^{(\mathrm{i})}(\pi_{XY})\nonumber \\
 & :=\mathcal{R}(\pi_{XY})\\
 & =\left\{ \begin{array}{rcl}
\left(R_{0},R\right) & : & \exists P_{W}P_{X|W}P_{Y|W}\textrm{ s.t. }\\
P_{XY} & = & \pi_{XY},\\
R & \ge & I(W;X),\\
R_{0}+R & \ge & -H(XY|W)+\sum_{w}P(w)\\
 &  & \times\mathcal{H}(P_{X|W=w},P_{Y|W=w}\|\pi_{XY})
\end{array}\right\} ,\label{eq:IB}
\end{align}
and 
\begin{align}
 & \mathcal{R}^{(\mathrm{o})}(\pi_{XY})\nonumber \\
 & :=\left\{ \begin{array}{rcl}
\left(R_{0},R\right) & : & \exists P_{W}P_{X|W}P_{Y|W}\textrm{ s.t. }\\
P_{XY} & = & \pi_{XY},\\
R & \ge & I(W;X),\\
R_{0}+R & \ge & \Gamma\left(P_{W}P_{X|W}P_{Y|W},\pi_{XY}\right)
\end{array}\right\} .\label{eq:OB}
\end{align}
where 
\begin{align}
 & \Gamma\left(P_{W}P_{X|W}P_{Y|W},\pi_{XY}\right)\nonumber \\
 & :=-H(XY|W)+\underset{Q_{WW'}\in C(P_{W},P_{W})}{\min}\sum_{w,w'}Q(w,w')\nonumber \\
 & \qquad\times\mathcal{H}(P_{X|W=w},P_{Y|W=w'}\|\pi_{XY}).
\end{align}
For $\left(X,Y\right)$ with finite alphabets, Cuff \cite{Cuff} showed
that $\mathcal{R}_{\mathrm{Cuff}}(\pi_{XY})$ is equal to the admissible
rate region for the TV-approximate channel synthesis problem (see
the definition in Subsection \ref{subsec:TV-approximate-Channel-Synthesis}).
For \eqref{eq:-TV} and \eqref{eq:IB}, it suffices to restrict the
alphabet size of $W$ such that $|\mathcal{W}|\le|\mathcal{X}||\mathcal{Y}|+1$.

By utilizing the multi-letter expression in Theorem \ref{thm:multiletter},
we provide single-letter inner and outer bounds for the admissible
rate region. The proof of Theorem \ref{thm:singleletter} is given
in Appendix \ref{sec:singleletter}.
\begin{thm}[Single-letter Bounds]
\label{thm:singleletter} For a joint distribution $\pi_{XY}$ defined
on a finite alphabet, 
\begin{align}
 & \mathcal{R}^{(\mathrm{i})}(\pi_{XY})\subseteq\mathcal{R}_{\mathrm{Exact}}(\pi_{XY})\subseteq\mathcal{R}^{(\mathrm{o})}(\pi_{XY})\cap\mathcal{R}_{\mathrm{Cuff}}(\pi_{XY}).
\end{align}
\end{thm}
\begin{rem}
Note that the only difference between the inner bound $\mathcal{R}^{(\mathrm{i})}(\pi_{XY})$
and outer bound $\mathcal{R}_{\mathrm{Cuff}}(\pi_{XY})$ is that in
$\mathcal{R}_{\mathrm{Cuff}}(\pi_{XY})$, the sum-rate is lower bounded
by $I\left(W;XY\right)$, but in $\mathcal{R}^{(\mathrm{i})}(\pi_{XY})$,
the sum-rate is lower bounded by $-H(XY|W)+\sum_{w}P(w)\mathcal{H}(P_{X|W=w},P_{Y|W=w}\|\pi_{XY})$.
By the definition of maximal cross-entropy, 
\begin{align}
 & \sum_{w}P(w)\mathcal{H}(P_{X|W=w},P_{Y|W=w}\|\pi_{XY})\nonumber \\
 & \geq\sum_{w}P(w)\sum_{x,y}P_{X|W}\left(x|w\right)P_{Y|W}\left(y|w\right)\log\frac{1}{\pi\left(x,y\right)}\\
 & =H\left(\pi_{XY}\right).
\end{align}
Hence we have 
\begin{align}
 & -H(XY|W)+\sum_{w}P(w)\mathcal{H}(P_{X|W=w},P_{Y|W=w}\|\pi_{XY})\nonumber \\
 & \geq I\left(W;XY\right).
\end{align}
That is, our lower bound on the sum rate is at least as large as Cuff's;
see \eqref{eq:-TV}.
\end{rem}
\begin{rem}
The only difference between the inner bound $\mathcal{R}^{(\mathrm{i})}(\pi_{XY})$
and outer bound $\mathcal{R}^{(\mathrm{o})}(\pi_{XY})$ is that in
$\mathcal{R}^{(\mathrm{o})}(\pi_{XY})$, the minimization is taken
over all couplings of $\left(P_{W},P_{W}\right)$, but in $\mathcal{R}^{(\mathrm{i})}(\pi_{XY})$,
it is not (or equivalently, the expectation in \eqref{eq:IB} can
be seen as being taken under the equality coupling of $\left(P_{W},P_{W}\right)$,
namely $P_{W}(w)1\{w'=w\}$).
\end{rem}
\begin{rem}
It can be shown that the sum rate bound in $\mathcal{R}^{(\mathrm{o})}(\pi_{XY})$
\begin{align}
 & -H(XY|W)+\min_{Q_{WW'}\in C(P_{W},P_{W})}\sum_{w,w'}Q_{WW'}(w,w')\nonumber \\
 & \qquad\times\mathcal{H}(P_{X|W=w},P_{Y|W=w'}\|\pi_{XY})\label{eq:-34}\\
 & =\min_{Q_{WW'}\in C(P_{W},P_{W})}\sum_{w,w'}Q_{WW'}(w,w')\max_{\substack{Q_{XY}\in\\
C(P_{X|W=w},P_{Y|W=w'})
}
}\nonumber \\
 & \qquad\sum_{x,y}Q(x,y)\log\frac{P_{X|W}(x|w)P_{Y|W}(y|w')}{\pi\left(x,y\right)}\\
 & \geq\min_{Q_{WW'}\in C(P_{W},P_{W})}\sum_{w,w'}Q_{WW'}(w,w')\nonumber \\
 & \qquad\times D\left(P_{X|W=w}P_{Y|W=w'}\|\pi_{XY}\right)\\
 & \geq0.
\end{align}
Moreover, there exists some distribution $\pi_{XY}$ such that $\mathcal{R}^{(\mathrm{o})}(\pi_{XY})\subsetneqq\mathcal{R}_{\mathrm{Cuff}}(\pi_{XY})$
(e.g., the DSBS; see the next subsection). However, it is difficult
to compare \eqref{eq:-34} to $I(XY;W)$ for all $\pi_{XY}$. As yet,
we have not found any distribution $\pi_{XY}$ such that $\mathcal{R}^{(\mathrm{o})}(\pi_{XY})\backslash\mathcal{R}_{\mathrm{Cuff}}(\pi_{XY})\neq\emptyset$.
\end{rem}
Define 
\begin{align}
R^{*}\left(R_{0}\right) & :=\inf_{\left(R,R_{0}\right)\in\mathcal{R}_{\mathrm{Exact}}(\pi_{XY})}R\\
R_{0}^{*}\left(R\right) & :=\inf_{\left(R,R_{0}\right)\in\mathcal{R}_{\mathrm{Exact}}(\pi_{XY})}R_{0}.\label{eq:-14}
\end{align}
Then from the inner and outer bounds in Theorem \ref{thm:singleletter},
we have that 
\begin{equation}
R^{*}\left(\infty\right)=I_{\pi}(X;Y).
\end{equation}
This is consistent with Bennett et al.'s observation \cite[Theorem 2]{bennett2002entanglement}.
That is, when there exists unlimited shared randomness available at
the encoder and decoder, a target channel can be synthesized by some
scheme if and only if the minimum asymptotic communication rate is
larger than or equal to the mutual information $I_{\pi}(X;Y)$ between
$(X,Y)\sim\pi_{XY}$. Moreover, the authors of \cite{bennett2002entanglement}
also showed that an exponential number of bits (infinite rate) of
shared randomness suffices to realize such synthesis. Bennett et al.{}
\cite[pp.~2939]{bennett2014quantum} conjectured that when communication
rates are restricted to approach to the optimal communication rate
$I_{\pi}(X;Y)$ asymptotically as $n\to\infty$, an exponential amount
of shared randomness is also necessary{} to realize exact synthesis.
(As mentioned in the introduction, this is referred to as the BDHSW
conjecture). Harsha et al.{} \cite{harsha2010communication} (as
well as Li and El Gamal \cite{li2018strong}) disproved this conjecture
for discrete $(X,Y)$ with finite alphabets, and showed that for this
case shared randomness with rate $\log|\mathcal{Y}|$ is sufficient
to realize such synthesis. Now we also focus on this limiting case
(in which the asymptotic optimal communication rate is used), and
investigate the minimum amount of shared randomness required for this
case. We have the following upper bound. 
\begin{align}
 & R_{0}^{*}\left(I_{\pi}(X;Y)\right)\nonumber \\
 & =\inf_{\left(I_{\pi}(X;Y),R_{0}\right)\in\mathcal{R}_{\mathrm{Exact}}(\pi_{XY})}R_{0}\\
 & \leq\min_{P_{W|Y}:X\rightarrow W\rightarrow Y}-H(X)-H(Y|W)\nonumber \\
 & \qquad+\sum_{w}P(w)\mathcal{H}(P_{X|W=w},P_{Y|W=w}\|\pi_{XY})\label{eq:-31}\\
 & \leq H_{\pi}(Y|X),\label{eq:-32}
\end{align}
where \eqref{eq:-31} follows from the inner bound in Theorem \ref{thm:singleletter},
and \eqref{eq:-32} follows by setting $W=Y$. The upper bound is,
in general, tighter than Harsha et al.'s bound of $\log|\mathcal{Y}|$.
In the next subsection, we will show that the upper bound in \eqref{eq:-32}
is tight for the DSBS. In Bennett et al.'s scheme \cite{bennett2014quantum},
the shared randomness is used to generate a random codebook. However,
in our scheme, as described after Theorem \ref{thm:multiletter},
we apply a mixture decomposition technique to construct a variable-length
exact synthesis code, which is a mixture of a fixed-length $\infty$-R\'enyi-approximate
code and a completely lossless code. As mentioned previously, the
performance of our scheme primarily depends on the performance of
the $\infty$-R\'enyi-approximate code which requires a much lower rate
of shared randomness than Bennett et al.'s scheme. Furthermore, in
Harsha et al.'s code \cite{harsha2010communication} (as well as Li
and El Gamal's \cite{li2018strong}), it is required that $Y^{n}$
and $\left(X^{n},K_{n}\right)$ are related as $Y^{n}=f\left(X^{n},K_{n}\right)$
for some deterministic function $f$. However, in our scheme, such
a requirement is not necessary. Hence it is not unexpected that our
bound is tighter than those presented in \cite{harsha2010communication,li2018strong}
.

We also have the following lower bound. 
\begin{align}
 & R_{0}^{*}\left(I_{\pi}(X;Y)\right)\nonumber \\
 & \geq\min_{P_{W|Y}:X\rightarrow W\rightarrow Y}-H(X)-H(Y|W)\\
 & \qquad+\min_{Q_{WW'}\in C(P_{W},P_{W})}\sum_{w,w'}Q_{WW'}(w,w')\nonumber \\
 & \qquad\times\mathcal{H}(P_{X|W=w},P_{Y|W=w'}\|\pi_{XY}),\label{eq:-31-2}
\end{align}
where \eqref{eq:-31-2} follows from the outer bound in Theorem \ref{thm:singleletter}.
Note that $\mathcal{R}_{\mathrm{Cuff}}(\pi_{XY})$ is the admissible
rate region for the TV-approximate synthesis problem \cite{Cuff10},
and it is also an outer bound on $\mathcal{R}_{\mathrm{Exact}}(\pi_{XY})$.
On the other hand, for TV-approximate synthesis, the corresponding
minimum shared randomness rate $\widetilde{R}_{0}^{*}\left(I_{\pi}(X;Y)\right):=\inf_{\left(I_{\pi}(X;Y),R_{0}\right)\in\mathcal{R}_{\mathrm{Cuff}}(\pi_{XY})}R_{0}$
is equal to the {\em necessary conditional entropy} \cite{Cuff10}
\begin{equation}
H_{\pi}(Y\dagger X):=\min_{f:X\rightarrow f(Y)\rightarrow Y}H(f(Y)|X).
\end{equation}
Hence, $R_{0}^{*}\left(I_{\pi}(X;Y)\right)$ is also lower bounded
by $H_{\pi}(Y\dagger X)$.

\subsection{\label{subsec:Doubly-Symmetric-Binary}Doubly Symmetric Binary Source}

A DSBS is a source $\left(X,Y\right)$ with distribution 
\begin{equation}
\pi_{XY}:=\left[\begin{array}{cc}
\alpha_{0} & \beta_{0}\\
\beta_{0} & \alpha_{0}
\end{array}\right]\label{eq:-10}
\end{equation}
where $\alpha_{0}=\frac{1-p}{2},\beta_{0}=\frac{p}{2}$ with $p\in[0,\frac{1}{2}]$.
This is equivalent to $X\sim\mathrm{Bern}(\frac{1}{2})$ and $Y=X\oplus E$
with $E\sim\mathrm{Bern}(p)$ independent of $X$; or $X=W\oplus A$
and $Y=W\oplus B$ with $W\sim\mathrm{Bern}(\frac{1}{2})$, $A\sim\mathrm{Bern}(a)$,
and $B\sim\mathrm{Bern}(b)$ mutually independent, where $a\in(0,p),a\overline{b}+\overline{a}b=p$.
Here w.l.o.g., we restrict $p\in[0,\frac{1}{2}]$, since otherwise,
we can set $X\oplus1$ to $X$.

By utilizing the inner and outer bounds in Theorem \ref{thm:singleletter},
we completely characterize the admissible rate region for the DSBS.
The proof of Theorem \ref{thm:DSBS} is given in Appendix \ref{sec:DSBS}. 
\begin{thm}
\label{thm:DSBS}For a DSBS $\left(X,Y\right)$ with distribution
$\pi_{XY}$ given in \eqref{eq:-10}, 
\begin{align}
 & \mathcal{R}_{\mathrm{Exact}}(\pi_{XY})\nonumber \\
 & =\left\{ \begin{array}{rcl}
\left(R,R_{0}\right) & : & a\in[0,p],b:=\frac{p-a}{1-2a},\\
R & \ge & 1-H_{2}(a),\\
R_{0}+R & \ge & \log\frac{1}{\alpha_{0}}+\left(a+b\right)\log\frac{\alpha_{0}}{\beta_{0}}\\
 &  & -H_{2}(a)-H_{2}(b)
\end{array}\right\} ,\label{eq:-19}
\end{align}
where $H_{2}(x):=-x\log x-(1-x)\log(1-x)$ denotes the binary entropy
function. 
\end{thm}
Following steps similar to the proof of Theorem \ref{thm:DSBS}, one
can show that for the DSBS, the admissible rate region for TV-approximate
channel synthesis is
\begin{align}
 & \mathcal{R}_{\mathrm{TV}}(\pi_{XY})\nonumber \\
 & =\left\{ \begin{array}{rcl}
\left(R,R_{0}\right) & : & a\in[0,p],b:=\frac{p-a}{1-2a},\\
R & \ge & 1-H_{2}(a),\\
R_{0}+R & \ge & 1+H_{2}(p)-H_{2}(a)-H_{2}(b)
\end{array}\right\} .\label{eq:-21}
\end{align}
Obviously, $\mathcal{R}_{\mathrm{Exact}}(\pi_{XY})$ and $\mathcal{R}_{\mathrm{TV}}(\pi_{XY})$
only differ in the final inequalities. It is easy to check that for
$p\in(0,\frac{1}{2})$, $\mathcal{R}_{\mathrm{Exact}}(\pi_{XY})$
is strictly included in $\mathcal{R}_{\mathrm{TV}}(\pi_{XY})$. Intuitively,
this interesting consequence is caused by the type overflow phenomenon,
which was described in detail after Theorem \ref{thm:multiletter}.
Such a result also confirms that type overflow does not affect TV-approximate
synthesis, but it does affect exact synthesis. 

Furthermore, letting $R=1-H_{2}(p)$ in $\mathcal{R}_{\mathrm{Exact}}(\pi_{XY})$,
we get $a=p,b=0$. Hence $R_{0}\ge H_{2}(p)$. That is, the upper
bound in \eqref{eq:-32} is tight for the DSBS.

\begin{figure*}
\centering \includegraphics[width=0.7\textwidth]{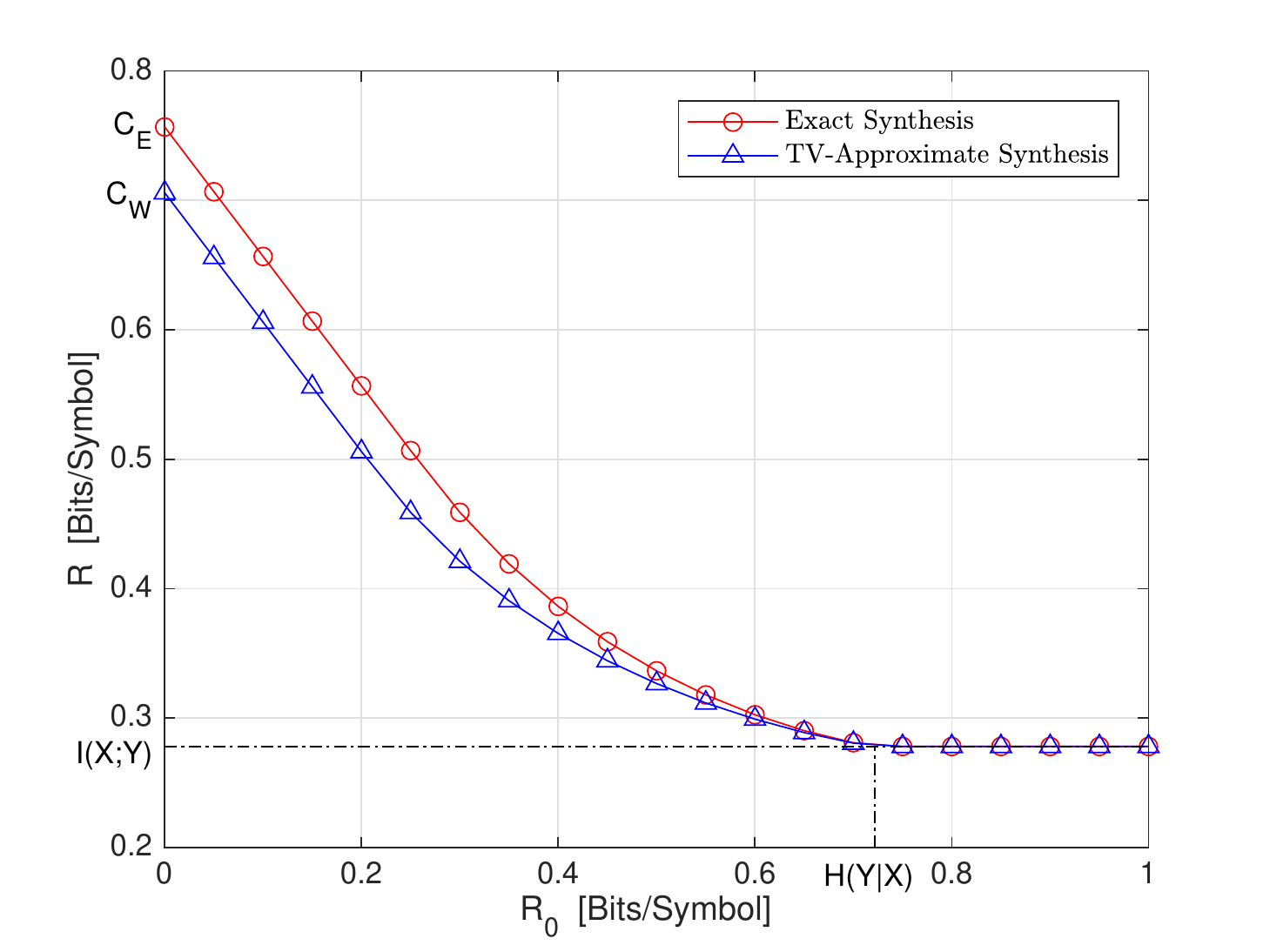}

\caption{\label{fig:Common-informations-for}Illustrations of the admissible
regions for exact synthesis (given in \eqref{eq:-19}) and TV-approximate
synthesis (given in \eqref{eq:-21}) for the DSBS $(X,Y)$ in which
$X\sim\mathrm{Bern}(\frac{1}{2})$ and $Y=X\oplus E$ with $E\sim\mathrm{Bern}(0.2)$
(i.e., $p=0.2$) independent of $X$. The admissible regions for exact
and TV-approximate syntheses respectively correspond to the regions
above the two curves. In this figure, $C_{\mathrm{W}}$ and $C_{\mathrm{E}}$
respectively denote Wyner's \cite{Wyner} and the exact common informations
\cite{Kumar,yu2018on} between $(X,Y)$, which also respectively correspond
to the minimum $R$'s for TV-approximate and exact syntheses when
$R_{0}=0$.}
\end{figure*}

The admissible regions for exact synthesis and TV-approximate synthesis
for the DSBS are illustrated in Fig.~\ref{fig:Common-informations-for}.

\section{Extension to More General Distributions}

\subsection{\label{subsec:TV-approximate-Channel-Synthesis}TV-Approximate Channel
Synthesis}

We extend the TV-approximate channel synthesis problem to the case
of more general distributions. In contrast to the exact synthesis
problem, in the TV-approximate synthesis problem, the communication
rate is measured by the exponent of the alphabet size of $W_{n}$,
rather than the normalized conditional entropy of $W_{n}$ given $K_{n}$.
In addition, the generated (or synthesized) distribution $\pi_{X}^{n}P_{Y^{n}|X^{n}}$
is required to approach $\pi_{XY}^{n}$ asymptotically under the TV
distance, instead of being exactly equal to $\pi_{XY}^{n}$.
\begin{defn}
A fixed-length $(n,R_{0},R)$-code consists of a pair of random mappings
$P_{W_{n}|X^{n}K_{n}}:\mathcal{X}^{n}\times\calK_{n}\to{\cal W}_{n},P_{Y^{n}|W_{n}K_{n}}:{\cal W}_{n}\times\calK_{n}\to\mathcal{Y}^{n}$
for some countable set ${\cal W}_{n}$ such that $\frac{1}{n}\log\left|{\cal W}_{n}\right|\le R$.
\end{defn}
\begin{defn}
The admissible region of shared randomness rate and communication
rate for the TV-approximate channel synthesis problem is defined as
\begin{align}
 & \mathcal{R}_{\mathrm{TV}}(\pi_{XY}):=\nonumber \\
 & \mathrm{cl}\left\{ \begin{array}{l}
(R_{0},R):\exists\left\{ \textrm{fixed-length }(n,R_{0},R)\textrm{ code}\right\} _{n=1}^{\infty}\textrm{ s.t.}\\
\qquad\lim_{n\to\infty}\left|\pi_{X}^{n}P_{Y^{n}|X^{n}}-\pi_{XY}^{n}\right|=0
\end{array}\right\} .
\end{align}
\end{defn}
For $\pi_{XY}$ with a finite alphabet, $\mathcal{R}_{\mathrm{TV}}(\pi_{XY})$
was completely characterized by Cuff \cite{Cuff} and Bennett et al.
\cite{bennett2014quantum}. For an arbitrary distribution $\pi_{XY}$
defined on an arbitrary measurable space, the inner bound in the following
theorem was proven by Cuff \cite[Theorem II.1]{Cuff} (since as mentioned
by the author, \cite[Theorem~II.1]{Cuff} also holds for ``general''
distributions), and the outer bound in the following theorem follows
by Cuff's converse proof of \cite[Theorem~II.1]{Cuff}.
\begin{thm}[TV-approximate Channel Synthesis]
\cite{Cuff} \label{thm:GeneralWyner} For a source $(X,Y)$ with
an arbitrary distribution $\pi_{XY}$ (defined on the product of two
arbitrary measurable spaces), 
\begin{equation}
\mathcal{R}_{\mathrm{Cuff}}(\pi_{XY})\subseteq\mathcal{R}_{\mathrm{TV}}(\pi_{XY})\subseteq\widetilde{\mathcal{R}}_{\mathrm{Cuff}}(\pi_{XY}),\label{eqn:Wyner}
\end{equation}
where\footnote{For an arbitrary distribution $P_{WX}$, the mutual information $I(W;X)$
exists if $P_{WX}\ll P_{W}P_{X}$ and the integral $\int_{\mathcal{W}\times\mathcal{X}}\left|\log\frac{\mathrm{d}P_{WX}}{\mathrm{d}\left(P_{W}P_{X}\right)}\right|\mathrm{d}P_{WX}<+\infty$.
The mutual information always exists for distributions with finite
alphabets but does not always exist for other distributions. Hence
for an arbitrary distribution $\pi_{XY}$, we modify the definition
of $\mathcal{R}_{\mathrm{Cuff}}(\pi_{XY})$ given in \eqref{eq:-TV}
to the one given in \eqref{eq:-TV-1}. For brevity, we use the same
notation for these two definitions. This is consistent since the $\mathcal{R}_{\mathrm{Cuff}}(\pi_{XY})$
defined in \eqref{eq:-TV-1} reduces to the one defined in \eqref{eq:-TV}
when the distribution $\pi_{XY}$ has a finite alphabet (by standard
cardinality bounding techniques, for this case, it suffices to consider
distributions $P_{W}P_{X|W}P_{Y|W}$ with finite alphabets).} 
\begin{align}
 & \mathcal{R}_{\mathrm{Cuff}}(\pi_{XY})\nonumber \\
 & :=\left\{ \begin{array}{rcl}
\left(R_{0},R\right) & : & \exists P_{W}P_{X|W}P_{Y|W}\textrm{ s.t. }\\
P_{XY} & = & \pi_{XY},\\
I(W;X) & \textrm{ and} & I\left(W;XY\right)\textrm{ exist},\\
R & \ge & I(W;X),\\
R_{0}+R & \ge & I\left(W;XY\right)
\end{array}\right\} ,\label{eq:-TV-1}
\end{align}
and
\begin{align}
 & \widetilde{\mathcal{R}}_{\mathrm{Cuff}}(\pi_{XY})\nonumber \\
 & :=\lim_{\epsilon\downarrow0}\left\{ \begin{array}{rcl}
\left(R,R_{0}\right) & : & \exists P_{W}P_{X|W}P_{Y|W}\textrm{ s.t. }\\
\left|P_{XY}-\pi_{XY}\right| & \le & \epsilon,\\
I(W;X) & \textrm{ and} & I\left(W;XY\right)\textrm{ exist},\\
R & \ge & I(W;X),\\
R_{0}+R & \ge & I\left(W;XY\right)
\end{array}\right\} .
\end{align}
\end{thm}
Obviously, $\mathcal{R}_{\mathrm{Cuff}}(\pi_{XY})\subseteq\widetilde{\mathcal{R}}_{\mathrm{Cuff}}(\pi_{XY})$.
We do not know if they are equal in general. However,  for sources
with finite alphabets, $\mathcal{R}_{\mathrm{Cuff}}(\pi_{XY})=\widetilde{\mathcal{R}}_{\mathrm{Cuff}}(\pi_{XY})$
was proven by Cuff \cite{Cuff} and Bennett et al. \cite{bennett2014quantum}.
In the following, we show that $\mathcal{R}_{\mathrm{Cuff}}(\pi_{XY})=\widetilde{\mathcal{R}}_{\mathrm{Cuff}}(\pi_{XY})$
also holds for sources with countably infinite alphabets and some
class of continuous sources.
\begin{cor}
\label{cor:countable} Let $(X,Y)$ be a source with distribution
$\pi_{XY}$ defined on the product of two countably infinite alphabets.
Assume $H(\pi_{XY})$ exists (and hence is finite). Then we have
\begin{equation}
\mathcal{R}_{\mathrm{TV}}(\pi_{XY})=\mathcal{R}_{\mathrm{Cuff}}(\pi_{XY}).\label{eqn:stronger-1-2-1}
\end{equation}
\end{cor}
\begin{cor}
\label{cor:continuous} Assume $\pi_{XY}$ is an absolutely continuous
distribution on $\mathbb{R}^{2}$ such that its pdf\footnote{For brevity, we use the same notation $\pi_{XY}$ to denote both an
absolutely continuous distribution and the corresponding pdf. } $\pi_{XY}$ is log-concave\footnote{A pdf $\pi_{XY}$ is log-concave if $\log\pi_{XY}$ is concave.}
and differentiable. Assume $I(X;Y)$ exists (and hence is finite).
For $d>0$, define 
\begin{equation}
L_{d}:=\sup_{\left(x,y\right)\in[-d,d]^{2}}\frac{\left|\frac{\partial\pi_{XY}}{\partial x}\left(x,y\right)\right|+\left|\frac{\partial\pi_{XY}}{\partial y}\left(x,y\right)\right|}{\pi_{XY}\left(x,y\right)},\label{eq:-78}
\end{equation}
and 
\begin{equation}
\epsilon_{d}:=1-\pi_{XY}\left([-d,d]^{2}\right).\label{eq:-6}
\end{equation}
If there exists a function $\Delta\left(d\right)$ such that\footnote{Here $o_{+}\left(g(d)\right)$ denotes a positive function $f(d)$
such that $f(d)=o\left(g(d)\right)$ (i.e., ${\displaystyle \lim_{d\to\infty}{\frac{f(d)}{g(d)}}=0})$.} $\Delta\left(d\right)=de^{-o_{+}\left(\frac{1}{\epsilon_{d}}\right)}$
and $\Delta\left(d\right)=o_{+}\left(\left(dL_{d}\right)^{-\alpha}\right)$
for some $\alpha>1$ as $d\to\infty$, then 
\begin{equation}
\mathcal{R}_{\mathrm{TV}}(\pi_{XY})=\mathcal{R}_{\mathrm{Cuff}}(\pi_{XY}).\label{eqn:stronger-1-2-1-1}
\end{equation}
\end{cor}
\begin{IEEEproof}[Proofs of Corollary \ref{cor:countable} and Corollary \ref{cor:continuous}]
The regions $\mathcal{R}_{\mathrm{Cuff}}(\pi_{XY})$ and $\widetilde{\mathcal{R}}_{\mathrm{Cuff}}(\pi_{XY})$
are respectively characterized by the following functions 
\begin{align}
 & R_{\mathrm{Cuff}}^{*}\left(R_{0}\right)\nonumber \\
 & :=\inf_{\left(R,R_{0}\right)\in\mathcal{R}_{\mathrm{Cuff}}(\pi_{XY})}R\\
 & =\inf_{\substack{P_{W}P_{X|W}P_{Y|W}:\\
P_{XY}=\pi_{XY}
}
}\max\left\{ I\left(W;XY\right)-R_{0},I(W;X)\right\} ,\\
 & \widetilde{R}_{\mathrm{Cuff}}^{*}\left(R_{0}\right)\nonumber \\
 & :=\inf_{\left(R,R_{0}\right)\in\widetilde{\mathcal{R}}_{\mathrm{Cuff}}(\pi_{XY})}R\\
 & =\lim_{\epsilon\downarrow0}\inf_{\substack{P_{W}P_{X|W}P_{Y|W}:\\
\left|P_{XY}-\pi_{XY}\right|\le\epsilon
}
}\max\left\{ I\left(W;XY\right)-R_{0},I(W;X)\right\} .
\end{align}
Hence to prove that 
\begin{equation}
\widetilde{\mathcal{R}}_{\mathrm{Cuff}}(\pi_{XY})\subseteq\mathcal{R}_{\mathrm{Cuff}}(\pi_{XY}),\label{eqn:Wyner-1}
\end{equation}
we only need to show that 
\begin{equation}
R_{\mathrm{Cuff}}^{*}\left(R_{0}\right)\leq\widetilde{R}_{\mathrm{Cuff}}^{*}\left(R_{0}\right).\label{eqn:Wyner-1-1}
\end{equation}
In \cite[Corollaries 2 and 3]{yu2018on}, we applied techniques involving
truncation, mixture decomposition, discretization, and dyadic decomposition
(studied in \cite{li2017distributed}) to show that\footnote{More precisely, what we showed is the inequality \eqref{eq:-27} with
the constraint the TV-distance replaced by the relative entropy. However,
as mentioned in \cite[Theorem 5]{yu2018on}, the proofs of \cite[Corollaries 2 and 3]{yu2018on}
also apply to the TV-distance case. } 
\begin{align}
 & \inf_{P_{W}P_{X|W}P_{Y|W}:P_{XY}=\pi_{XY}}I\left(W;XY\right)\nonumber \\
 & \leq\lim_{\epsilon\downarrow0}\inf_{P_{W}P_{X|W}P_{Y|W}:\left|P_{XY}-\pi_{XY}\right|\le\epsilon}I\left(W;XY\right).\label{eq:-27}
\end{align}
The difference between \eqref{eqn:Wyner-1-1} and \eqref{eq:-27}
is only the objective functions. Following similar steps to the proofs
of \cite[Corollaries 2 and 3]{yu2018on}, one can show \eqref{eqn:Wyner-1-1}. 
\end{IEEEproof}
It is easy to verify that any bivariate Gaussian source with a correlation
coefficient $\rho\in(-1,1)$ satisfies the conditions given in Corollary
\ref{cor:continuous} \cite{yu2018on}. Hence we have the following
result. Without loss of any generality, we assume the correlation
coefficient $\rho$ between $(X,Y)$ is nonnegative; otherwise, we
can set $-X$ to $X$. 
\begin{cor}
\label{cor:Gaussian} For a Gaussian source $(X,Y)$ with correlation
coefficient $\rho\in[0,1),$ we have 
\begin{align}
 & \mathcal{R}_{\mathrm{TV,G}}(\pi_{XY})=\mathcal{R}_{\mathrm{Cuff}}(\pi_{XY})\label{eq:Gaussian-1}\\
 & =\left\{ \begin{array}{rcl}
\left(R,R_{0}\right) & : & \alpha\in[\rho,1],\alpha\beta=\rho,\\
R & \ge & \frac{1}{2}\log\left[\frac{1}{1-\alpha^{2}}\right],\\
R_{0}+R & \ge & \frac{1}{2}\log\left[\frac{1-\rho^{2}}{\left(1-\alpha^{2}\right)\left(1-\beta^{2}\right)}\right]
\end{array}\right\} .\label{eq:Gaussian-1-1}
\end{align}
\end{cor}
\begin{IEEEproof}
The equality in \eqref{eq:Gaussian-1} follows from Theorem \ref{cor:continuous}
by verifying the assumption holds for Gaussian sources. The equality
in \eqref{eq:Gaussian-1-1} can be proven by a similar proof to that
of Wyner's common information of Gaussian sources \cite[Theorems 2 and  8]{xu2013wyner,yu2016generalized},
and hence the proof is omitted here. 
\end{IEEEproof}

\subsection{Exact Channel Synthesis}

\subsubsection{Discrete Channels with Countably Infinite Alphabets}

In the proof of Theorem \ref{thm:multiletter}, a truncated i.i.d.
code was adopted to prove the achievability part. In this ensemble,
codewords are i.i.d., each drawn according to a set of truncated distributions,
obtained by truncating a set of product distributions into some (strongly)
typical sets. For the countably infinite alphabet case, we replace
strongly typical sets with unified typical sets \cite{ho2010information,ho2010markov}.
Then we can establish the following result. The proof is omitted. 
\begin{cor}
\label{cor:countable-1-1} Let $(X,Y)$ be a source with distribution
$\pi_{XY}$ defined on the product of two countably infinite alphabets.
We have 
\begin{equation}
\widetilde{\mathcal{R}}^{(\mathrm{i})}(\pi_{XY})\subseteq\mathcal{R}_{\mathrm{Exact}}(\pi_{XY})\subseteq\widetilde{\mathcal{R}}^{(\mathrm{o})}(\pi_{XY})\cap\mathcal{R}_{\mathrm{Cuff}}(\pi_{XY}),\label{eqn:stronger-1-1-2-1-1}
\end{equation}
where $\mathcal{R}_{\mathrm{Cuff}}(\pi_{XY})$ was defined in \eqref{eq:-TV},
\begin{align}
 & \widetilde{\mathcal{R}}^{(\mathrm{i})}(\pi_{XY})\nonumber \\
 & :=\lim_{\epsilon\downarrow0}\left\{ \begin{array}{rcl}
\left(R_{0},R\right) & : & \exists P_{W}P_{X|W}P_{Y|W}\textrm{ s.t. }\\
P_{XY} & = & \pi_{XY},\\
R & \ge & I(W;X),\\
R_{0}+R & \ge & \Gamma_{1}\left(P_{W}P_{X|W}P_{Y|W},\pi_{XY}\right)
\end{array}\right\} ,\label{eq:UB-1}
\end{align}
and 
\begin{align}
 & \widetilde{\mathcal{R}}^{(\mathrm{o})}(\pi_{XY})\nonumber \\
 & :=\lim_{\epsilon\downarrow0}\left\{ \begin{array}{rcl}
\left(R,R_{0}\right) & : & \exists P_{W}P_{X|W}P_{Y|W}\textrm{ s.t. }\\
D\left(P_{XY}\|\pi_{XY}\right) & \le & \epsilon,\\
R & \ge & I(W;X),\\
R_{0}+R & \ge & \Gamma_{2}\left(P_{W}P_{X|W}P_{Y|W},\pi_{XY}\right)
\end{array}\right\} \label{eq:-59-1}
\end{align}
with 
\begin{align}
 & \Gamma_{1}\left(P_{W}P_{X|W}P_{Y|W},\pi_{XY}\right)\nonumber \\
 & :=-H(XY|W)+\underset{\substack{Q_{XYW}:D\left(Q_{WX}\|P_{WX}\right)\le\epsilon,\\
D\left(Q_{WY}\|P_{WY}\right)\le\epsilon
}
}{\sup}\sum_{w,x,y}P(w)\nonumber \\
 & \qquad\times Q\left(x,y|w\right)\log\frac{1}{\pi\left(x,y\right)}
\end{align}
and
\begin{align}
 & \Gamma_{2}\left(P_{W}P_{X|W}P_{Y|W},\pi_{XY}\right)\nonumber \\
 & :=-H(XY|W)+\underset{Q_{WW'}\in C(P_{W},P_{W})}{\inf}\sum_{w,w'}Q_{WW'}(w,w')\nonumber \\
 & \qquad\times\mathcal{H}(P_{X|W=w},P_{Y|W=w'}\|\pi_{XY}).
\end{align}
\end{cor}
For the finite alphabet case, since $\mathcal{P}(\mathcal{W}\times\mathcal{X}\times\mathcal{Y})$
is compact, we can take $\epsilon=0$ in both \eqref{eq:UB-1} and
\eqref{eq:-59-1}. This follows by the Bolzano--Weierstrass theorem,
i.e., passing to a convergent subsequence of distributions. However,
for the countably infinite alphabet case, in general we cannot do
this. Furthermore, it may be possible to remove both $\epsilon$'s
in the optimizations in \eqref{eq:UB-1} and \eqref{eq:-59-1} by
truncating the distributions, transforming them into finite alphabet
ones, as in the proof of Corollary \cite[Corollary 2]{yu2018on}.
This will be studied in the future.

\subsubsection{Gaussian Sources}

Before considering Gaussian sources, we first consider more general
continuous sources which satisfy certain regularity conditions. For
such sources, we prove a sufficient condition for exact channel synthesis,
which is analogous to Lemma \ref{lem:sufficiency}. For $\epsilon>0$,
define the truncated distribution 
\begin{equation}
\widetilde{\pi}_{X^{n}}(x^{n}):=\frac{\pi_{X}^{n}(x^{n})1\{x^{n}\in\mathcal{A}_{\epsilon}^{\left(n\right)}\}}{\pi_{X}^{n}(\mathcal{A}_{\epsilon}^{\left(n\right)})}\textrm{ for }n\ge1.\label{eq:-73-5}
\end{equation}
 
\begin{lem}
\label{lem:continuous} Assume $\pi_{XY}$ is an absolutely continuous
distribution on $\mathbb{R}^{2}$ with $\mathbb{E}\left[X^{2}\right],\mathbb{E}\left[Y^{2}\right]<\infty$.
Without loss of generality, we assume $\mathbb{E}\left[X^{2}\right]=\mathbb{E}\left[Y^{2}\right]=1$.
Assume $I(X;Y)$ exists (and hence is finite). Assume for every $x$,
the pdf $\pi_{Y|X}\left(\cdot|x\right)$ is log-concave and continuously
differentiable. For $\epsilon>0$ and $n\in\mathbb{N}$, define 
\begin{equation}
L_{\epsilon,n}:=\sup_{\left(x,y\right)\in\mathcal{\mathcal{L}}_{\epsilon,n}^{2}}\left|\frac{\partial}{\partial y}\log\pi_{Y|X}\left(y|x\right)\right|,\label{eq:-97}
\end{equation}
where\footnote{In \eqref{eq:-36}, the reason why we choose the number $2$ in the
bound $\sqrt{n\left(2+\epsilon\right)}$ is because that from $\mathbb{E}\left[X^{2}\right]=\mathbb{E}\left[Y^{2}\right]=1$,
we have $\mathbb{E}\left[\left(Y-X\right)^{2}\right]\leq2$.}{} 
\begin{equation}
\mathcal{\mathcal{L}}_{\epsilon,n}:=\left\{ x\in\mathbb{R}:|x|\leq\sqrt{n\left(2+\epsilon\right)}\right\} .\label{eq:-36}
\end{equation}
Assume $\log L_{\epsilon,n}$ is sub-exponential in $n$ for fixed
$\epsilon$. Then if there exists a sequence of fixed-length codes
with rates $\left(R_{0},R\right)$ that generates a channel $P_{Y^{n}|X^{n}}$
for the source $X^{n}\sim\widetilde{\pi}_{X^{n}}$ such that 
\begin{equation}
D_{\infty}(P_{Y^{n}|X^{n}}\|\pi_{Y|X}^{n}|\widetilde{\pi}_{X^{n}})=o\left(\frac{1}{n+\log L_{\epsilon,n}}\right)\label{eq:-28-2}
\end{equation}
for any $\epsilon>0$, then there exists a sequence of variable-length
codes with rates $\left(R_{0},R\right)$ that exactly generates $\pi_{XY}^{n}$
(i.e., the sequence of codes generates a channel $\pi_{Y|X}^{n}$
for the source $X^{n}\sim\pi_{X}^{n}$).
\end{lem}
\begin{rem}
\label{rem:Gaussian}One important example satisfying the conditions
in the lemma above is the bivariate Gaussian source. Consider a bivariate
Gaussian source $\pi_{XY}=\mathcal{N}\left(0,\Sigma_{XY}\right)$
where $\Sigma_{XY}=\left[\begin{array}{cc}
1 & \rho\\
\rho & 1
\end{array}\right]$ with $\rho\in[0,1)$. That is, 
\begin{equation}
Y=\rho X+\sqrt{1-\rho^{2}}Z,
\end{equation}
where $Z\sim\mathcal{N}\left(0,1\right)$ is independent of $X$.
For this case, 
\begin{align}
L_{\epsilon,n} & =\sup_{\left(x,y\right)\in\mathcal{\mathcal{L}}_{\epsilon,n}^{2}}\left|\frac{y-\rho x}{1-\rho^{2}}\right|=\frac{\sqrt{n\left(2+\epsilon\right)}}{1-\rho}.
\end{align}
Hence $\log L_{\epsilon,n}$ is sub-exponential in $n$ for fixed
$\epsilon$. Observe that $\frac{1}{n+\log L_{\epsilon,n}}\sim\frac{1}{n}$.
Hence if there exists a sequence of fixed-length codes with rates
$\left(R_{0},R\right)$ that generates $P_{Y^{n}|X^{n}}$ such that
\begin{equation}
D_{\infty}(P_{Y^{n}|X^{n}}\|\pi_{Y|X}^{n}|\widetilde{\pi}_{X^{n}})=o\left(\frac{1}{n}\right),\label{eq:-29}
\end{equation}
then there must exist a sequence of variable-length codes with rates
$\left(R_{0},R\right)$ that exactly generates $\pi_{XY}^{n}$.  
\end{rem}
\begin{IEEEproof}[Proof of Lemma \ref{lem:continuous}]
 The proof follows similar ideas as the one of \cite[Lemma 2]{yu2018on},
where the techniques of mixture decomposition, truncation, discretization,
and Li and El Gamal's dyadic decomposition \cite{li2017distributed}
were used. However, the difference is that in \cite[Lemma 2]{yu2018on},
we used the mixture decomposition technique to decompose a memoryless
correlated source{} $\pi_{XY}^{n}$, but here we need to decompose
a memoryless channel{} $\pi_{Y|X}^{n}$. To address this problem,
we need to combine the proof of \cite[Lemma 2]{yu2018on} with the
one of Lemma \ref{lem:sufficiency}. We omit the details of the proof
for the sake of brevity.
\end{IEEEproof}
Next we prove an inner bound on $\mathcal{R}_{\mathrm{Exact}}(\pi_{XY})$
for Gaussian sources $\pi_{XY}$. Without loss of generality, we assume
that the correlation coefficient $\rho\in[0,1)$ between $(X,Y)$
is nonnegative; otherwise, we can set $-X$ as $X$. By substituting
$X=\alpha W+A,Y=\beta W+B,$ $P_{W}=\mathcal{N}(0,1),P_{X|W}(\cdot|w)=\mathcal{N}(w,1-\alpha^{2}),P_{Y|W}(\cdot|w)=\mathcal{N}(w,1-\beta^{2})$
into the inner bound \eqref{eq:IB}, we obtain the following inner
bound for Gaussian sources. Although the inner bound \eqref{eq:IB}
is shown for sources with finite alphabets, one can prove an analogous
inner bound for the Gaussian case. 
\begin{thm}
\label{thm:Gaussian} For a Gaussian source $(X,Y)$ with correlation
coefficient $\rho\in[0,1)$, we have 
\begin{equation}
\widetilde{\mathcal{R}}^{(\mathrm{i})}(\pi_{XY})\subseteq\mathcal{R}_{\mathrm{Exact}}(\pi_{XY})\subseteq\mathcal{R}_{\mathrm{TV,G}}(\pi_{XY}),\label{eq:Gaussian}
\end{equation}
where $\mathcal{R}_{\mathrm{TV,G}}(\pi_{XY})$ is given in \eqref{eq:Gaussian-1},
and 
\begin{align}
 & \widetilde{\mathcal{R}}^{(\mathrm{i})}(\pi_{XY})\nonumber \\
 & =\left\{ \begin{array}{rcl}
\left(R,R_{0}\right) & : & \alpha\in[\rho,1],\alpha\beta=\rho,\\
R & \ge & \frac{1}{2}\log\left[\frac{1}{1-\alpha^{2}}\right],\\
R_{0}+R & \ge & \frac{1}{2}\log\left[\frac{1-\rho^{2}}{\left(1-\alpha^{2}\right)\left(1-\beta^{2}\right)}\right]\\
 &  & +\frac{\rho\sqrt{\left(1-\alpha^{2}\right)\left(1-\beta^{2}\right)}}{1-\rho^{2}}
\end{array}\right\} .\label{eq:-18}
\end{align}
\end{thm}
\begin{IEEEproof}
Since Gaussian sources satisfy the regularity conditions given in
Lemma \ref{lem:continuous} (see Remark \ref{rem:Gaussian}), by Lemma
\ref{lem:continuous} we only need to show that there exists a sequence
of fixed-length codes with rates $\left(R_{0},R\right)$ that generates
$P_{Y^{n}|X^{n}}$ satisfying \eqref{eq:-29}. This can be proven
by following similar steps to the achievability proof of Theorem \ref{thm:multiletter}
(which is for the discrete case); see Appendix \ref{subsec:achievability}.
In Appendix \ref{subsec:achievability}, a distributed version and
a centralized version of R\'enyi-covering lemmas are applied. For Gaussian
sources, the distributed R\'enyi-covering lemma was proven in \cite[Lemma 13]{yu2018on}.
Hence we only need to prove a centralized R\'enyi-covering lemma for
Gaussian sources. To this end, we combine the proof of \cite[Lemma 13]{yu2018on}
with that of Lemma \ref{lem:Renyicovering-1}. We omit the details
of the proof. 
\end{IEEEproof}
The difference between the inner bound $\widetilde{\mathcal{R}}^{(\mathrm{i})}(\pi_{XY})$
for exact Gaussian synthesis and the admissible region $\mathcal{R}_{\mathrm{TV,G}}(\pi_{XY})$
for TV-approximate Gaussian synthesis is similar to that for the DSBS
case; see the discussion at the end of Subsection \ref{subsec:Doubly-Symmetric-Binary}.
Furthermore, for the DSBS, our inner bound in Theorem \ref{thm:DSBS}
is tight. Hence by the type overflow argument, we conjecture that
for Gaussian sources, the inner bound in \eqref{eq:Gaussian} is also
tight.

For a Gaussian source $(X,Y)$ with correlation coefficient $\rho\in[0,1),$
the mutual information $I_{\pi}(X;Y)=\frac{1}{2}\log\left[\frac{1}{1-\rho^{2}}\right]$.
Hence for this case, under the condition that communication rates
are restricted to approach to $I_{\pi}(X;Y)$ asymptotically, the
minimum asymptotic rates of shared randomness required for exact synthesis
and TV-approximate synthesis satisfy 
\begin{align}
R_{0}^{*}\left(I_{\pi}(X;Y)\right) & \geq\inf_{\left(I_{\pi}(X;Y),R_{0}\right)\in\mathcal{R}_{\mathrm{TV,G}}(\pi_{XY})}R_{0}\label{eq:-31-1-2-1}\\
 & =\infty.\label{eq:-32-1-1-1}
\end{align}
Hence for a Gaussian source $(X,Y)$, any finite rate of shared randomness
cannot{} realize exact synthesis or TV-approximate synthesis when
there is no penalty on the asymptotic communication rate.

\begin{figure*}
\centering \includegraphics[width=0.7\textwidth]{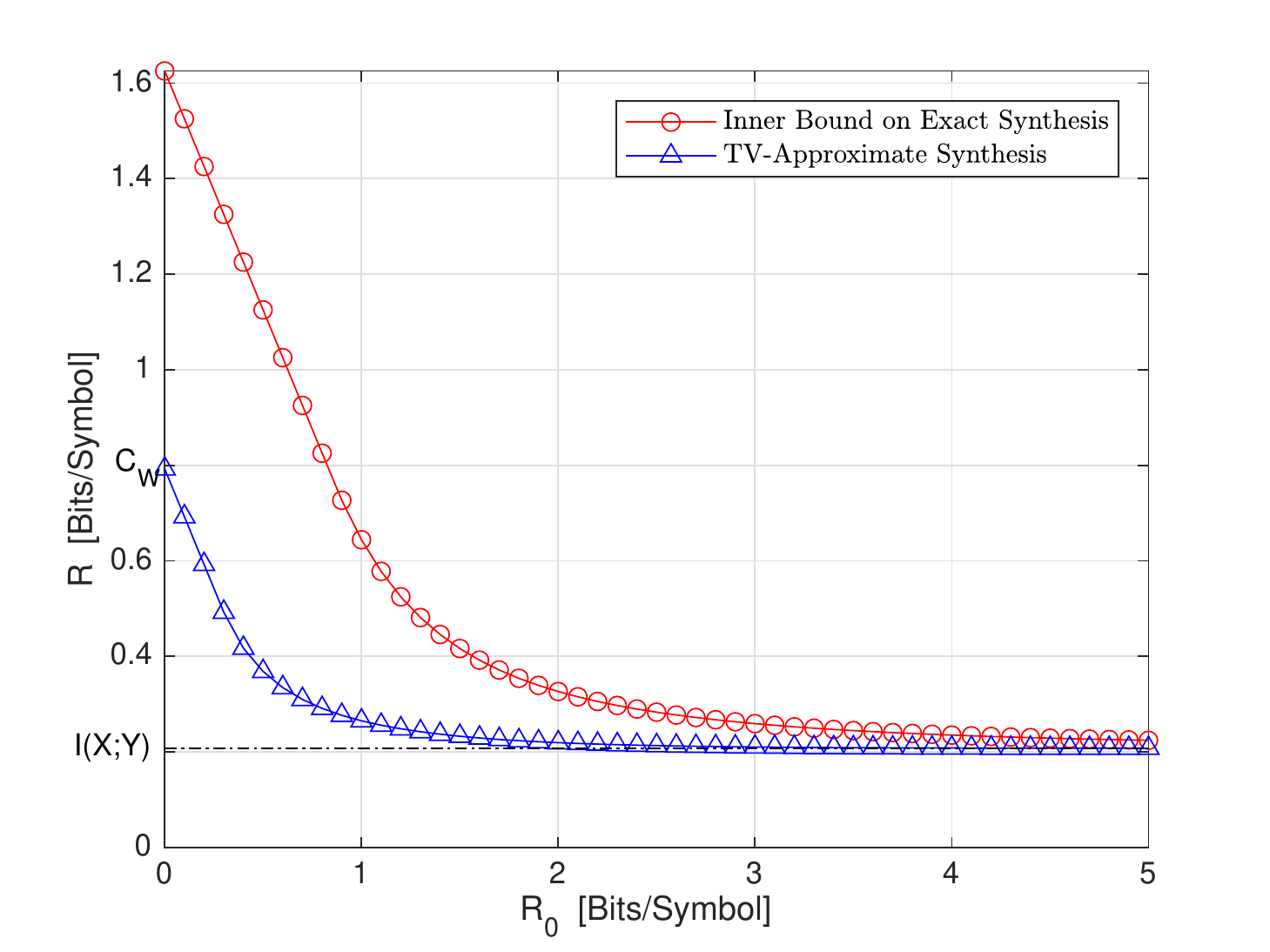}

\caption{\label{fig:Common-informations-for-1}Illustrations of our inner bound
$\widetilde{\mathcal{R}}^{(\mathrm{i})}(\pi_{XY})$ in \eqref{eq:-18}
for exact synthesis and the admissible region $\mathcal{R}_{\mathrm{TV,G}}(\pi_{XY})$
in \eqref{eq:Gaussian-1} for TV-approximate synthesis for Gaussian
sources with correlation coefficient $\rho=0.5$. The admissible region
for TV-approximate synthesis corresponds to the region above the curve
with markers ``$\triangle$''. The boundary of the admissible region
for exact synthesis lies between between the two curves in the figure.}
\end{figure*}

For Gaussian sources, our inner bound $\widetilde{\mathcal{R}}^{(\mathrm{i})}(\pi_{XY})$
in \eqref{eq:-18} for exact synthesis and the admissible region $\mathcal{R}_{\mathrm{TV,G}}(\pi_{XY})$
in \eqref{eq:Gaussian-1} for TV-approximate synthesis are illustrated
in Fig.~\ref{fig:Common-informations-for-1}. The boundary of $\mathcal{R}_{\mathrm{Exact}}(\pi_{XY})$
lies between between the two curves in the figure. 

\section{Concluding Remarks}

\label{sec:concl}In this paper, we studied the tradeoff between the
shared randomness rate and the communication rate for exact channel
synthesis; provided single-letter inner and outer bounds on the admissible
rate region for this problem; completely characterized them for the
DSBS; and extended these results, and also existing results for TV-approximate
channel synthesis, to discrete sources with countably infinite alphabets
and continuous sources (including Gaussian sources). For the DSBS,
we observed that the admissible rate region for exact channel synthesis
is strictly larger than that for TV-approximate channel synthesis.
For Gaussian sources with correlation coefficient $\rho\in[0,1)$,
we provided inner and outer bounds on the admissible rate region for
exact channel synthesis.{} Due to the intuitive type overflow argument,
we conjecture that the inner bound is tight.

Exact and TV-approximate channel synthesis differ primarily in the
following two aspects. First, for TV-approximate synthesis, it does
not matter whether one uses a fixed- or variable-length message. In
contrast, for exact synthesis, such a distinction is consequential.
This is because by the asymptotic equipartition property, i.i.d. copies
of a variable-length (nonuniform) message are distributed over a typical
set with high probability. Hence by using superblock coding (in which
block codes are considered as supersymbols \cite{Cover}) and truncation
techniques \cite{loeve1977elementary,cramer2016mathematical}, a code
for TV-approximate synthesis with a variable-length message can be
converted into a code for TV-approximate synthesis with a fixed-length
message, with the approximation error increasing by an asymptotically
vanishing amount. On the other hand, a fixed-length message can be
seen as a special case of variable-length message. Hence variable-length
TV-approximate synthesis and the fixed-length version are equivalent.
That is, they have the same admissible rate region. However, this
is not true for the exact synthesis problem. In general, for the exact
case, using a fixed-length message will result in a strictly larger
admissible rate region. The conversion argument above does not apply
to the exact synthesis problem, since increasing the approximation
error by an asymptotically vanishing value is not allowed for exact
synthesis. Hence by comparing (variable-length) exact synthesis and
variable-length TV-approximate synthesis, we know that exact synthesis
is strictly more difficult to realize than the TV-approximate version.
This leads to requiring larger rates required for exact synthesis.
Second, the type overflow phenomenon does not affect TV-approximate
synthesis, but indeed plays a critical role for $\infty$-R\'enyi-approximate
synthesis (or exact synthesis). In distributed simulation problems,
random variables are required to form Markov chains. Hence type overflow
is unavoidable (at least for the DSBS). Truncated i.i.d. coding is
an effective approach to mitigate the effect of type overflow. Hence
it is useful for exact synthesis, but not for TV-approximate synthesis.

One may expect our synthesis scheme to be useful in analyzing zero-error
coding problems. In this class of problems, the channel to be synthesized
is the identity channel. However, our synthesis scheme does not seem
to be useful in this case. To enable our synthesis scheme to be applicable
to zero-error coding, it is required to find a code with synthesized
channel $P_{Y^{n}|X^{n}}$ such that 
\begin{align}
D_{\infty}(P_{Y^{n}|X^{n}}\|\pi_{Y|X}^{n}|\widetilde{\pi}_{X^{n}})\to0,
\end{align}
where $\widetilde{\pi}_{X^{n}}$ is a truncated version of $\pi_{X}^{n}$.
For the identity channel $\pi_{Y|X}$, this requirement is too strong
or restrictive. In fact, by the definition of $D_{\infty}$, it is
equivalent to the stringent condition that $P_{Y^{n}|X^{n}}=\pi_{Y|X}^{n}$
for every sufficiently large $n$. Therefore, such conversion from
the exact synthesis problem to the $\infty$-R\'enyi-approximate synthesis
problem does not appear to help us to relax the requirement of exact
synthesis.


\appendices{}

\section{\label{sec:entropycharacterization}Proof of Proposition \ref{prop:entropycharacterization}}

By using \eqref{eq:-17}, we have that 
\begin{align}
 & \mathcal{R}_{\mathrm{Exact}}(\pi_{XY})\nonumber \\
 & =\left\{ \begin{array}{l}
(R_{0},R):\exists\left\{ \left(P_{W_{n}|X^{n}K_{n}},P_{Y^{n}|W_{n}K_{n}}\right)\right\} _{n=1}^{\infty}\textrm{ s.t. }\\
\qquad P_{Y^{n}|X^{n}}=\pi_{Y|X}^{n},\forall n\\
\qquad R\ge\limsup_{n\to\infty}\frac{1}{n}H(W_{n}|K_{n})
\end{array}\right\} .
\end{align}

Define 
\begin{align}
a & :=\inf_{\substack{\left\{ \left(P_{W_{n}|X^{n}K_{n}},P_{Y^{n}|W_{n}K_{n}}\right)\right\} _{n=1}^{\infty}:\\
P_{Y^{n}|X^{n}}=\pi_{Y|X}^{n},\forall n
}
}\limsup_{n\to\infty}\frac{1}{n}H(W_{n}|K_{n})\label{eq:-41}\\
b & :=\limsup_{n\to\infty}\inf_{\substack{\left(P_{W_{n}|X^{n}K_{n}},P_{Y^{n}|W_{n}K_{n}}\right):\\
P_{Y^{n}|X^{n}}=\pi_{Y|X}^{n}
}
}\frac{1}{n}H(W_{n}|K_{n}).
\end{align}
Now we claim that 
\begin{equation}
a=b.\label{eq:-42}
\end{equation}
This is because by definition, for any $\epsilon>0$, there exists
a sequence of codes $\left\{ \left(Q_{W_{n}|X^{n}K_{n}},Q_{Y^{n}|W_{n}K_{n}}\right)\right\} _{n=1}^{\infty}$
such that $Q_{Y^{n}|X^{n}}=\pi_{Y|X}^{n}$ for all $n\ge1$ and $\limsup_{n\to\infty}\frac{1}{n}H_{Q}(W_{n}|K_{n})\leq a+\epsilon$.
Since in definition of $b$, we minimize $\frac{1}{n}H(W_{n}|K_{n})$
for each $n$, we have 
\begin{equation}
\limsup_{n\to\infty}\frac{1}{n}H_{Q}(W_{n}|K_{n})\geq b.
\end{equation}
Therefore, $b\le a+\epsilon$. Since $\epsilon>0$ is arbitrary, we
have $b\le a$.

On the other hand, by the definition of $b$, for any $\epsilon>0$,
there exists an integer $N$ such that for all $n\ge N$, 
\begin{equation}
\inf_{\substack{\left(P_{W_{n}|X^{n}K_{n}},P_{Y^{n}|W_{n}K_{n}}\right):\\
P_{Y^{n}|X^{n}}=\pi_{Y|X}^{n}
}
}\frac{1}{n}H(W_{n}|K_{n})\leq b+\epsilon.
\end{equation}
This implies that for each $n\ge N$, there exists an $n$-length
code $\left(Q_{W_{n}|X^{n}K_{n}},Q_{Y^{n}|W_{n}K_{n}}\right)$ such
that $Q_{Y^{n}|X^{n}}=\pi_{Y|X}^{n}$ and $\frac{1}{n}H_{Q}(W_{n}|K_{n})\leq b+2\epsilon$.
Such a sequence of codes satisfies the constraint in \eqref{eq:-41}.
Hence $a\le\limsup_{n\to\infty}\frac{1}{n}H_{Q}(W_{n}|K_{n})\leq b+2\epsilon$.
Since $\epsilon>0$ is arbitrary, we have $b\ge a$.

Combining the two points above, we have $a=b$. Therefore, we obtain
that 
\begin{align}
 & \mathcal{R}_{\mathrm{Exact}}(\pi_{XY})\nonumber \\
 & =\left\{ (R_{0},R):R\ge a\right\} \label{eq:-15}\\
 & =\left\{ (R_{0},R):R\ge b\right\} \label{eq:-30-1-1-1}\\
 & =\left\{ \begin{array}{l}
(R_{0},R):R\ge\\
\quad\lim_{n\to\infty}\inf_{\substack{\left(P_{W_{n}|X^{n}K_{n}},P_{Y^{n}|W_{n}K_{n}}\right):\\
P_{Y^{n}|X^{n}}=\pi_{Y|X}^{n}
}
}\frac{1}{n}H(W_{n}|K_{n})
\end{array}\right\} \label{eq:-24}\\
 & =\left\{ \begin{array}{l}
(R_{0},R):R\ge\\
\quad\inf_{n\ge1}\inf_{\substack{\left(P_{W_{n}|X^{n}K_{n}},P_{Y^{n}|W_{n}K_{n}}\right):\\
P_{Y^{n}|X^{n}}=\pi_{Y|X}^{n}
}
}\frac{1}{n}H(W_{n}|K_{n})
\end{array}\right\} \label{eq:-25}\\
 & =\mathrm{cl}\,\bigcup_{n\ge1}\left\{ \begin{array}{l}
(R_{0},R):R\ge\\
\quad\inf_{\substack{\left(P_{W_{n}|X^{n}K_{n}},P_{Y^{n}|W_{n}K_{n}}\right):\\
P_{Y^{n}|X^{n}}=\pi_{Y|X}^{n}
}
}\frac{1}{n}H(W_{n}|K_{n})
\end{array}\right\} \label{eq:-43}\\
 & =\mathrm{cl}\,\bigcup_{n\ge1}\left\{ \begin{array}{l}
(R_{0},R):\exists\left(P_{W_{n}|X^{n}K_{n}},P_{Y^{n}|W_{n}K_{n}}\right)\textrm{ s.t. }\\
\qquad P_{Y^{n}|X^{n}}=\pi_{Y|X}^{n},\\
\qquad R\ge\frac{1}{n}H(W_{n}|K_{n})
\end{array}\right\} ,\label{eq:-15-1-1}
\end{align}
where  \eqref{eq:-24} follows from Fekete's subadditive lemma since
$a_{n}:=\inf_{\left(P_{W_{n}|X^{n}K_{n}},P_{Y^{n}|W_{n}K_{n}}\right):P_{Y^{n}|X^{n}}=\pi_{Y|X}^{n}}H(W_{n}|K_{n})$
is a subadditive sequence (i.e., $a_{m+n}\le a_{m}+a_{n}$); \eqref{eq:-25}
follows since the sequence of infima is non-increasing; \eqref{eq:-43}
follows from the definition of the union and infimizations operations;
and \eqref{eq:-15-1-1} follows similarly to \eqref{eq:-42}.

\section{\label{sec:crossentropy}Proof of Proposition \ref{prop:maximalcrossentropy}}

Obviously, both \eqref{eq:maximalcrossentropy-1} and \eqref{eq:maximalcrossentropy-1-1-1}
follow directly from the definition of the maximal cross-entropy.
Next we consider necessary and sufficient conditions for which the
equalities hold. We first consider \eqref{eq:maximalcrossentropy-1}.
\begin{align}
 & \mathcal{H}(\pi_{X},\pi_{Y}\|\pi_{XY})-H(\pi_{XY})\nonumber \\
 & =\max_{P_{XY}\in C(\pi_{X},\pi_{Y})}\sum_{x,y}\left(P(x,y)-\pi\left(x,y\right)\right)\nonumber \\
 & \qquad\times\log\frac{1}{\pi\left(x,y\right)}\\
 & \geq\sum_{x,y}\left(\pi(x)\pi(y)-\pi\left(x,y\right)\right)\log\frac{1}{\pi\left(x,y\right)}\\
 & =\sum_{x,y}\left(\pi(x)\pi(y)-\pi\left(x,y\right)\right)\log\frac{\pi(x)\pi(y)}{\pi\left(x,y\right)}\label{eq:-1}\\
 & =D\left(\pi_{X}\pi_{Y}\|\pi_{XY}\right)+D\left(\pi_{XY}\|\pi_{X}\pi_{Y}\right)\\
 & \geq0,
\end{align}
where \eqref{eq:-1} follows since $\sum_{x,y}\left(\pi(x)\pi(y)-\pi\left(x,y\right)\right)f(x)=0$
for any function $f(x)$. All the equalities above hold if and only
if $\pi_{XY}=\pi_{X}\pi_{Y}$.

We next consider \eqref{eq:maximalcrossentropy-1-1-1}. If $\pi_{XY}=\pi_{X}\pi_{Y}$,
then equality in \eqref{eq:maximalcrossentropy-1-1-1} holds. Next
we prove the ``only if'' part.

Assume $\mathcal{H}(P_{X},P_{Y}\|\pi_{XY})=\sum_{x,y}P_{X}(x)P_{Y}(y)\log\frac{1}{\pi\left(x,y\right)}$.
Obviously, $P_{X}P_{Y}\in C(P_{X},P_{Y})$. We add a perturbation
$\epsilon\eta(x,y)$ to the coupling $P_{X}P_{Y}$, where 
\begin{align}
\eta(x,y) & :=1\left\{ x=x_{0},y=y_{0}\right\} +1\left\{ x=x_{1},y=y_{1}\right\} \nonumber \\
 & \quad-1\left\{ x=x_{0},y=y_{1}\right\} -1\left\{ x=x_{1},y=y_{0}\right\} 
\end{align}
for some $x_{0},x_{1}\in\mathcal{X},y_{0},y_{1}\in\mathcal{Y}$. By
assumption, $\supp(\pi_{XY})=\supp(P_{X})\times\supp(P_{Y})$. Hence
for sufficiently small $\epsilon\in\mathbb{R}$, $P_{XY}^{(\epsilon)}:=P_{X}P_{Y}+\epsilon\eta$
is a distribution. Obviously, $P_{XY}^{(\epsilon)}\in\in C(P_{X},P_{Y})$.
Then we have 
\begin{align}
 & \mathcal{H}(P_{X},P_{Y}\|\pi_{XY})\nonumber \\
 & \geq\sum_{x,y}P_{XY}^{(\epsilon)}(x,y)\log\frac{1}{\pi\left(x,y\right)}\\
 & =\sum_{x,y}P_{X}(x)P_{Y}(y)\log\frac{1}{\pi\left(x,y\right)}+\sum_{x,y}\epsilon\eta(x,y)\log\frac{1}{\pi\left(x,y\right)}\\
 & =\sum_{x,y}P_{X}(x)P_{Y}(y)\log\frac{1}{\pi\left(x,y\right)}+\epsilon\log\frac{\pi\left(x_{0},y_{1}\right)\pi\left(x_{1},y_{0}\right)}{\pi\left(x_{0},y_{0}\right)\pi\left(x_{1},y_{1}\right)}.\label{eq:-7}
\end{align}
Hence 
\begin{equation}
\log\frac{\pi\left(x_{0},y_{1}\right)\pi\left(x_{1},y_{0}\right)}{\pi\left(x_{0},y_{0}\right)\pi\left(x_{1},y_{1}\right)}=0,\label{eq:-22}
\end{equation}
otherwise we can choose $\epsilon$ such that \eqref{eq:-7} is strictly
larger than $0$, which in turn implies $\mathcal{H}(P_{X},P_{Y}\|\pi_{XY})>\sum_{x,y}P_{X}(x)P_{Y}(y)\log\frac{1}{\pi\left(x,y\right)}$.
Since \eqref{eq:-22} holds for all $x_{0},x_{1}\in\mathcal{X},y_{0},y_{1}\in\mathcal{Y}$,
by simple algebraic calculations, we have 
\begin{equation}
\frac{\pi\left(x_{0},y_{1}\right)}{\pi\left(x_{0},y_{0}\right)}=\frac{\pi\left(y_{1}\right)}{\pi\left(y_{0}\right)},\label{eq:-22-2}
\end{equation}
and further obtain that 
\begin{equation}
\frac{\pi\left(x_{0}\right)-\pi\left(x_{0},y_{0}\right)}{\pi\left(x_{0},y_{0}\right)}=\frac{1-\pi\left(y_{0}\right)}{\pi\left(y_{0}\right)},\label{eq:-22-2-1}
\end{equation}
i.e., $\pi\left(x_{0},y_{0}\right)=\pi\left(x_{0}\right)\pi\left(y_{0}\right)$
for all $\left(x_{0},y_{0}\right)\in\mathcal{X}\times\mathcal{Y}$.
Hence $\pi_{XY}=\pi_{X}\pi_{Y}$.

\section{\label{sec:equivalence}Proof of Theorem \ref{thm:multiletter}}

\subsection{\label{subsec:achievability} Achievability }

Fix $\epsilon>0$ and define 
\begin{equation}
\widetilde{\pi}_{X^{n}}(x^{n}):=\frac{\pi_{X}^{n}(x^{n})1\{x^{n}\in\mathcal{T}_{\epsilon}^{\left(n\right)}\}}{\pi_{X}^{n}(\mathcal{T}_{\epsilon}^{\left(n\right)})},\textrm{ for }n\ge1.\label{eq:-73}
\end{equation}
To show the desired result, we need the following sufficiency result
for the exact synthesis, which states that the channel synthesis under
the $\infty$-R\'enyi divergence measure implies the exact channel synthesis. 
\begin{lem}
\label{lem:sufficiency}If there exists a sequence of fixed-length
codes with rates $\left(R_{0},R\right)$ that generates a channel
$P_{Y^{n}|X^{n}}$ for a source $X^{n}\sim\widetilde{\pi}_{X^{n}}$
such that 
\begin{equation}
D_{\infty}(P_{Y^{n}|X^{n}}\|\pi_{Y|X}^{n}|\widetilde{\pi}_{X^{n}})\to0,\label{eq:-28}
\end{equation}
then there exists a sequence of variable-length codes with rates $\left(R_{0},R\right)$
that exactly generates $\pi_{XY}^{n}$ (i.e., generates a channel
$\pi_{Y|X}^{n}$ for a source $X^{n}\sim\pi_{X}^{n}$).
\end{lem}
\begin{rem}
Here ``fixed-length codes'' in the condition can be relaxed to ``variable-length
codes''.
\end{rem}
\begin{rem}
Since $\pi_{X}^{n}(\mathcal{T}_{\epsilon}^{\left(n\right)})\to1$
as $n\to\infty$, the condition in \eqref{eq:-28} is equivalent to
that 
\begin{equation}
D_{\infty}(\widetilde{\pi}_{X^{n}}P_{Y^{n}|X^{n}}\|\pi_{XY}^{n})\to0.
\end{equation}
\end{rem}
\begin{IEEEproof}
We apply a technique so-called mixture decomposition to prove Lemma
\ref{lem:sufficiency}, which was previously used in \cite{ho2010interplay,vellambi2016sufficient,vellambi2018new}.
According to the definition of $D_{\infty}$, $D_{\infty}(P_{Y^{n}|X^{n}}\|\pi_{Y|X}^{n}|\widetilde{\pi}_{X^{n}})\leq\delta$
implies that $P_{Y^{n}|X^{n}}\left(y^{n}|x^{n}\right)\leq e^{\delta}\pi_{Y|X}^{n}\left(y^{n}|x^{n}\right)$
for all $\left(x^{n},y^{n}\right)\in\mathcal{T}_{\epsilon}^{\left(n\right)}\times\mathcal{Y}^{n}$.
For $x^{n}\in\mathcal{T}_{\epsilon}^{\left(n\right)}$, define 
\begin{equation}
\widehat{P}_{Y^{n}|X^{n}}\left(y^{n}|x^{n}\right):=\frac{e^{\delta}\pi_{Y|X}^{n}\left(y^{n}|x^{n}\right)-P_{Y^{n}|X^{n}}\left(y^{n}|x^{n}\right)}{e^{\delta}-1},
\end{equation}
then obviously, $\widehat{P}_{Y^{n}|X^{n}}$ is a conditional distribution.
Hence $\pi_{Y|X}^{n}$ can be written as a mixture distribution 
\begin{align}
 & \pi_{Y|X}^{n}\left(y^{n}|x^{n}\right)\nonumber \\
 & =e^{-\delta}P_{Y^{n}|X^{n}}\left(y^{n}|x^{n}\right)+\left(1-e^{-\delta}\right)\widehat{P}_{Y^{n}|X^{n}}\left(y^{n}|x^{n}\right)
\end{align}
for $x^{n}\in\mathcal{T}_{\epsilon}^{\left(n\right)}$. The encoder
first generates a Bernoulli random variable $U$ with $P_{U}(1)=e^{-\delta}$,
compresses it using $1$ bit, and transmits it to the two generators.
If $U=1$ and $x^{n}\in\mathcal{T}_{\epsilon}^{\left(n\right)}$,
then the encoder and decoder use the synthesis codes (prescribed in
the lemma) with rate $R$ to generate $P_{Y^{n}|X^{n}}$. If $U=0$
and $x^{n}\in\mathcal{T}_{\epsilon}^{\left(n\right)}$, then the encoder
generates $Y^{n}|X^{n}=x^{n}\sim\widehat{P}_{Y^{n}|X^{n}}\left(\cdot|x^{n}\right)$,
and uses a variable-length compression code with rate $\leq\log|\mathcal{Y}|$
to transmit $Y^{n}$. If $x^{n}\notin\mathcal{T}_{\epsilon}^{\left(n\right)}$,
then the encoder generates $Y^{n}|X^{n}=x^{n}\sim\pi_{Y|X}^{n}\left(\cdot|x^{n}\right)$,
and uses a variable-length compression code with rate $\leq\log|\mathcal{Y}|$
to transmit $Y^{n}$. The conditional distribution generated by such
a mixture code is $e^{-\delta}P_{Y^{n}|X^{n}}\left(y^{n}|x^{n}\right)+\left(1-e^{-\delta}\right)\widehat{P}_{Y^{n}|X^{n}}\left(y^{n}|x^{n}\right)$
for $x^{n}\in\mathcal{T}_{\epsilon}^{\left(n\right)}$ and $\pi_{Y|X}^{n}\left(y^{n}|x^{n}\right)$
for $x^{n}\notin\mathcal{T}_{\epsilon}^{\left(n\right)}$, i.e., $\pi_{Y|X}^{n}\left(y^{n}|x^{n}\right)$
for all $x^{n}$. The total communication rate is no larger than 
\begin{align}
 & \pi_{X}^{n}(\mathcal{T}_{\epsilon}^{\left(n\right)})\left(\frac{1}{n}+e^{-\delta}R+\left(1-e^{-\delta}\right)\log|\mathcal{Y}|\right)\nonumber \\
 & \qquad+\left(1-\pi_{X}^{n}(\mathcal{T}_{\epsilon}^{\left(n\right)})\right)\log|\mathcal{Y}|,
\end{align}
which converges to $R$ upon taking the limit in $n\to\infty$ and
the limit in $\delta\to0$. Furthermore, the rate of shared randomness
for this mixed code is still $R_{0}$. 
\end{IEEEproof}
By Lemma \ref{lem:sufficiency}, to show the achievability part, we
only need to show that there exists $\epsilon>0$ and a sequence of
synthesis codes with rates $\left(R_{0},R\right)$ that generates
$P_{Y^{n}|X^{n}}$ such that $D_{\infty}(\widetilde{\pi}_{X^{n}}P_{Y^{n}|X^{n}}\|\pi_{XY}^{n})\to0$.
Next we prove this.

To show the achievability part, we only need to show that the single-letter
expression $\mathcal{R}(\pi_{XY})$ satisfies $\mathcal{R}(\pi_{XY})\subseteq\mathcal{R}_{\mathrm{Exact}}(\pi_{XY})$.
This is because we can obtain the inner bound $\frac{1}{n}\mathcal{R}(\pi_{XY}^{n})$
by substituting $\pi_{XY}\leftarrow\pi_{XY}^{n}$ into the single-letter
expression.

Assume $Q_{WXY}$ is a distribution such that $Q_{WXY}=Q_{W}Q_{X|W}Q_{Y|W}$.
Similarly as in \eqref{eq:-73}, we define the distributions\footnote{Here $f(x)\propto g(x)$ denotes that $f(x)$ is proportional to $g(x)$,
i.e., $f(x)=c\cdot g(x)$ for some constant $c$.} 
\begin{align}
\widetilde{Q}_{W^{n}}\left(w^{n}\right) & \propto Q_{W}^{n}\left(w^{n}\right)1\left\{ w^{n}\in\mathcal{T}_{2\epsilon}^{\left(n\right)}\left(Q_{W}\right)\right\} ,\label{eq:-23}\\
\widetilde{Q}_{X^{n}|W^{n}}\left(x^{n}|w^{n}\right) & \propto Q_{X|W}^{n}\left(x^{n}|w^{n}\right)\nonumber \\
 & \times1\left\{ x^{n}\in\mathcal{T}_{4\epsilon}^{\left(n\right)}\left(Q_{WX}|w^{n}\right)\right\} ,\\
\widetilde{Q}_{Y^{n}|W^{n}}\left(y^{n}|w^{n}\right) & \propto Q_{Y|W}^{n}\left(y^{n}|w^{n}\right)\nonumber \\
 & \times1\left\{ y^{n}\in\mathcal{T}_{4\epsilon}^{\left(n\right)}\left(Q_{WY}|w^{n}\right)\right\} .\label{eq:-26}
\end{align}
We consider a random codebook $\mathcal{C}_{n}=\left\{ W^{n}\left(m,k\right)\right\} $
with $W^{n}\left(m,k\right),\left(m,k\right)\in\calM_{n}\times\calK_{n}$
drawn independently for different $\left(m,k\right)$'s and according
to the same distribution $\widetilde{Q}_{W^{n}}$. Define $P_{K_{n}}:=\mathrm{Unif}[1:e^{nR_{0}}]$,
$P_{M_{n}}:=\mathrm{Unif}[1:e^{nR}]$. For random mappings $\widetilde{Q}_{X^{n}|W^{n}}$
and $\widetilde{Q}_{Y^{n}|W^{n}}$, we define 
\begin{align}
 & \widehat{Q}_{X^{n}Y^{n}|\mathcal{C}_{n}}(x^{n},y^{n}|\left\{ W^{n}\left(m,k\right)\right\} )\nonumber \\
 & :=\sum_{k,m}P_{K_{n}}(k)P_{M_{n}}(m)\widetilde{Q}_{X^{n}|W^{n}}\left(x^{n}|W^{n}\left(m,k\right)\right)\nonumber \\
 & \qquad\times\widetilde{Q}_{Y^{n}|W^{n}}\left(y^{n}|W^{n}\left(m,k\right)\right),
\end{align}
which is the output distribution induced by the codebook $\mathcal{C}_{n}$
in a distributed source simulation system with simulators $\left(\widetilde{Q}_{X^{n}|W^{n}},\widetilde{Q}_{Y^{n}|W^{n}}\right)$.
For such a distribution, we have following two R\'enyi-covering lemmas
(i.e., soft-covering lemmas under R\'enyi divergence measures). Lemma
\ref{lem:Renyicovering} is proven in \cite[Lemma 6]{yu2018on}, and
the proof of Lemma \ref{lem:Renyicovering-1} is provided in Appendix
\ref{subsec:randomcode}. 
\begin{lem}[Distributed R\'enyi-Covering]
\cite[Lemma 6]{yu2018on} \label{lem:Renyicovering} For the random
code described above, if 
\begin{align}
R_{0}+R & >-H_{Q}(XY|W)\nonumber \\
 & +\sum_{w}Q(w)\mathcal{H}(Q_{X|W=w},Q_{Y|W=w}\|Q_{XY}),
\end{align}
then there exists some $\delta,\epsilon>0$ such that 
\begin{align}
 & \mathbb{P}_{\mathcal{C}_{n}}\left(D_{\infty}(\widehat{Q}_{X^{n}Y^{n}|\mathcal{C}_{n}}\|Q_{XY}^{n})\leq e^{-n\delta}\right)\to1
\end{align}
doubly exponentially fast. Here $\epsilon$ was used in the definitions
of $\widetilde{Q}_{W^{n}},\widetilde{Q}_{X^{n}|W^{n}}$, and $\widetilde{Q}_{Y^{n}|W^{n}}$
in \eqref{eq:-23}-\eqref{eq:-26}. 
\end{lem}
\begin{lem}[Centralized R\'enyi-Covering]
\label{lem:Renyicovering-1} Define a truncated product distribution
\begin{equation}
\widetilde{Q}_{X^{n}}(x^{n}):=\frac{Q_{X}^{n}(x^{n})1\{x^{n}\in\mathcal{T}_{\epsilon}^{\left(n\right)}\}}{Q_{X}^{n}(\mathcal{T}_{\epsilon}^{\left(n\right)})}.\label{eq:-73-2}
\end{equation}
For the random code described above, if 
\begin{equation}
R>I_{Q}(W;X),
\end{equation}
then there exists some $\delta,\epsilon>0$ such that 
\begin{align}
 & \mathbb{P}_{\mathcal{C}_{n}}\left(\begin{array}{c}
D_{\infty}(\widehat{Q}_{X^{n}|K_{n}\mathcal{C}_{n}}\|Q_{X}^{n}|P_{K_{n}})\leq e^{-n\delta},\\
D_{\infty}(\widetilde{Q}_{X^{n}}\|\widehat{Q}_{X^{n}|K_{n}\mathcal{C}_{n}}|P_{K_{n}})\leq e^{-n\delta}
\end{array}\right)\to1\label{eq:-11}
\end{align}
doubly exponentially fast. Here $\epsilon$ was used in the definitions
of $\widetilde{Q}_{X^{n}},\widetilde{Q}_{W^{n}},\widetilde{Q}_{X^{n}|W^{n}}$,
and $\widetilde{Q}_{Y^{n}|W^{n}}$. 
\end{lem}
\begin{rem}
\label{rem:If-,-then}If $R_{0}=0$, then \eqref{eq:-11} reduces
to the conclusion that 
\begin{align}
 & \mathbb{P}_{\mathcal{C}_{n}}\left(\begin{array}{c}
D_{\infty}(\widehat{Q}_{X^{n}|\mathcal{C}_{n}}\|Q_{X}^{n})\leq e^{-n\delta},\\
D_{\infty}(\widetilde{Q}_{X^{n}}\|\widehat{Q}_{X^{n}|\mathcal{C}_{n}})\leq e^{-n\delta}
\end{array}\right)\to1\label{eq:-12}
\end{align}
doubly exponentially fast. Since the convergence takes place doubly
exponentially fast and $K_{n}$ only can take exponential number of
different values, by the union bound, it can be seen that \eqref{eq:-11}
and \eqref{eq:-12} are equivalent.  
\end{rem}
\begin{rem}
Observe that, by definition, $\supp\left(\widetilde{Q}_{W^{n}}\widetilde{Q}_{X^{n}|W^{n}}\right)\subseteq\mathcal{T}_{4\epsilon}^{\left(n\right)}\left(Q_{WX}\right)$.
Hence for any codebook $c$, $\supp\left(\widehat{Q}_{X^{n}|\mathcal{C}_{n}=c}\right)\subseteq\mathcal{T}_{4\epsilon}^{\left(n\right)}\left(Q_{X}\right)$.
On the other hand, $Q_{X}^{n}(\mathcal{T}_{\epsilon}^{\left(n\right)}\left(Q_{X}\right))\to1$
exponentially fast. Hence Lemma \ref{lem:Renyicovering-1} implies
that there exists some $\delta',\epsilon>0$ such that with high probability
(probabilities approaching to $1$ doubly exponentially fast), the
codebook $\mathcal{C}_{n}$ satisfies that a) for any $x^{n}\in\mathcal{T}_{\epsilon}^{\left(n\right)}\left(Q_{X}\right)$,
\begin{equation}
\exp\left(-e^{-n\delta'}\right)\leq\frac{\widehat{Q}_{X^{n}|\mathcal{C}_{n}}\left(x^{n}\right)}{Q_{X}^{n}\left(x^{n}\right)}\leq\exp\left(e^{-n\delta'}\right);
\end{equation}
b) for any $x^{n}\in\mathcal{T}_{4\epsilon}^{\left(n\right)}\left(Q_{X}\right)\backslash\mathcal{T}_{\epsilon}^{\left(n\right)}\left(Q_{X}\right)$,
\begin{equation}
0\leq\frac{\widehat{Q}_{X^{n}|\mathcal{C}_{n}}\left(x^{n}\right)}{Q_{X}^{n}\left(x^{n}\right)}\leq\exp\left(e^{-n\delta'}\right);
\end{equation}
and c) for any $x^{n}\in\mathcal{X}^{n}\backslash\mathcal{T}_{4\epsilon}^{\left(n\right)}\left(Q_{X}\right)$,
\begin{equation}
\widehat{Q}_{X^{n}|\mathcal{C}_{n}}\left(x^{n}\right)=0.
\end{equation}
 
\end{rem}
\begin{rem}
It is worth noting that Lemma \ref{lem:Renyicovering} and Lemma \ref{lem:Renyicovering-1}
differ in two points. First, Lemma \ref{lem:Renyicovering} focuses
on a distributed{} setting where $X^{n}$ and $Y^{n}$ are respectively
generated by $W^{n}\left(M_{n},K_{n}\right)$ through truncated i.i.d.
channels $\widetilde{Q}_{X^{n}|W^{n}}$ and $\widetilde{Q}_{Y^{n}|W^{n}}$.
However, Lemma \ref{lem:Renyicovering-1} focuses on a centralized{}
setting where the output $X^{n}$ is generated by $W^{n}\left(M_{n},K_{n}\right)$
through only one truncated i.i.d. channel $\widetilde{Q}_{X^{n}|W^{n}}$.
Second, in Lemma \ref{lem:Renyicovering} only the $\infty$-R\'enyi
divergence between the real{} output distribution and the ideal{}
output distribution is considered; while in Lemma \ref{lem:Renyicovering-1}
both the $\infty$-R\'enyi divergence between the ideal{} output distribution
and the real{} output distribution and the $\infty$-R\'enyi divergence
between the real{} output distribution and the ideal{} output distribution
are considered.
\end{rem}
\begin{rem}
For the i.i.d. channel case (instead of the truncated version), R\'enyi-covering
lemmas were studied in \cite[Theorems 2-4]{yu2019renyi}.  
\end{rem}
Lemma \ref{lem:Renyicovering} and Lemma \ref{lem:Renyicovering-1}
imply that there exists $Q_{W}Q_{X|W}Q_{Y|W}$ with $Q_{XY}=\pi_{XY}$
such that for some $\delta,\epsilon>0$,
\begin{align}
 & \mathbb{P}_{\mathcal{C}_{n}}\left(\begin{array}{c}
D_{\infty}(\widehat{Q}_{X^{n}Y^{n}|\mathcal{C}_{n}}\|\pi_{XY}^{n})\leq e^{-n\delta},\\
D_{\infty}(\widetilde{\pi}_{X^{n}}\|\widehat{Q}_{X^{n}|K_{n}\mathcal{C}_{n}}|P_{K_{n}})\leq e^{-n\delta}
\end{array}\right)\to1
\end{align}
doubly exponentially fast, as long as the rate pair $\left(R_{0},R\right)$
is in the interior of $\mathcal{R}(\pi_{XY})$. Here $\epsilon$ was
used in the definition of $\widetilde{\pi}_{X^{n}}$; see \eqref{eq:-73}.
Hence there exists a sequence of deterministic codebooks $\left\{ c_{n}\right\} $
such that $D_{\infty}(\widehat{Q}_{X^{n}Y^{n}|\mathcal{C}_{n}=c_{n}}\|\pi_{XY}^{n})$
and $D_{\infty}(\widetilde{\pi}_{X^{n}}\|\widehat{Q}_{X^{n}|K_{n}\mathcal{C}_{n}=c_{n}}|P_{K_{n}})$
converge to zero exponentially fast. For such a sequence of deterministic
codebooks (under the condition $\mathcal{C}_{n}=c_{n}$), define 
\begin{align}
 & \widehat{Q}_{M_{n}K_{n}X^{n}Y^{n}}(m,k,x^{n},y^{n})\nonumber \\
 & :=P_{K_{n}}(k)P_{M_{n}}(m)\widetilde{Q}_{X^{n}|W^{n}}\left(x^{n}|w^{n}\left(m,k\right)\right)\nonumber \\
 & \qquad\times\widetilde{Q}_{Y^{n}|W^{n}}\left(y^{n}|w^{n}\left(m,k\right)\right)\\
 & =P_{K_{n}}\widehat{Q}_{X^{n}|K_{n}}\widehat{Q}_{M_{n}|X^{n}K_{n}}\widehat{Q}_{Y^{n}|M_{n}K_{n}}
\end{align}
and 
\begin{equation}
P_{M_{n}K_{n}X^{n}Y^{n}}:=P_{K_{n}}\widetilde{\pi}_{X^{n}}\widehat{Q}_{M_{n}|X^{n}K_{n}}\widehat{Q}_{Y^{n}|M_{n}K_{n}}.
\end{equation}
Now consider a synthesis code $\left(\widehat{Q}_{M_{n}|X^{n}K_{n}},\widehat{Q}_{Y^{n}|M_{n}K_{n}}\right)$.
Obviously, $P_{M_{n}K_{n}X^{n}Y^{n}}$ is the distribution induced
by such a synthesis code under the condition that the source $X^{n}\sim\widetilde{\pi}_{X^{n}}$.
Next we prove that such a synthesis code (with rates $\left(R_{0},R\right)$)
generates $P_{Y^{n}|X^{n}}$ such that $D_{\infty}(P_{X^{n}Y^{n}}\|\pi_{XY}^{n})\to0$.

Observe 
\begin{align}
D_{\infty}(P_{X^{n}Y^{n}}\|\pi_{XY}^{n}) & \leq D_{\infty}(P_{X^{n}Y^{n}}\|\widehat{Q}_{X^{n}Y^{n}})\nonumber \\
 & \qquad+D_{\infty}(\widehat{Q}_{X^{n}Y^{n}}\|\pi_{XY}^{n}).
\end{align}
By the choice of $\left\{ c_{n}\right\} $, the second term of the
right hand side above converges to zero exponentially. Next we consider
the first term. 
\begin{align}
 & D_{\infty}(P_{X^{n}Y^{n}}\|\widehat{Q}_{X^{n}Y^{n}})\nonumber \\
 & \leq D_{\infty}(P_{K_{n}}\widetilde{\pi}_{X^{n}}\widehat{Q}_{M_{n}|X^{n}K_{n}}\widehat{Q}_{Y^{n}|M_{n}K_{n}}\nonumber \\
 & \qquad\|P_{K_{n}}\widehat{Q}_{X^{n}|K_{n}}\widehat{Q}_{M_{n}|X^{n}K_{n}}\widehat{Q}_{Y^{n}|M_{n}K_{n}})\\
 & =D_{\infty}(\widetilde{\pi}_{X^{n}}\|\widehat{Q}_{X^{n}|K_{n}}|P_{K_{n}})\label{eq:-2}\\
 & \to0\textrm{ exponentially fast as }n\to\infty,\label{eq:-3}
\end{align}
where \eqref{eq:-3} follows by the choice of $\left\{ c_{n}\right\} $.
Hence 
\begin{align}
D_{\infty}(P_{X^{n}Y^{n}}\|\pi_{XY}^{n})= & D_{\infty}(\widetilde{\pi}_{X^{n}}P_{Y^{n}|X^{n}}\|\pi_{XY}^{n})\to0.
\end{align}
By Lemma \ref{lem:sufficiency}, we obtain the achievability part.

By definition, $\mathcal{R}_{\mathrm{Exact}}(\pi_{XY})$ is closed.
Hence $\mathrm{int}\mathcal{R}(\pi_{XY})\subseteq\mathcal{R}_{\mathrm{Exact}}(\pi_{XY})$
implies $\mathcal{R}(\pi_{XY})\subseteq\mathcal{R}_{\mathrm{Exact}}(\pi_{XY})$.

\subsubsection{\label{subsec:randomcode}Proof of Lemma \ref{lem:Renyicovering-1} }

Setting $Y$ to be a constant, Lemma \ref{lem:Renyicovering} implies
that there exists some $\delta>0$ such that\footnote{In fact, Lemma \ref{lem:Renyicovering} implies that for any $k\in[1:e^{nR_{0}}]$,
$\mathbb{P}_{\mathcal{C}_{n}}\left(D_{\infty}(\widehat{Q}_{X^{n}|\mathcal{C}_{n},K_{n}=k}\|Q_{X}^{n})\leq e^{-n\delta}\right)\to1$
doubly exponentially fast. However, as mentioned in Remark \ref{rem:If-,-then},
this is equivalent to \eqref{eq:-13}.} 
\begin{align}
 & \mathbb{P}_{\mathcal{C}_{n}}\left(D_{\infty}(\widehat{Q}_{X^{n}|K_{n}\mathcal{C}_{n}}\|Q_{X}^{n}|P_{K_{n}})\leq e^{-n\delta}\right)\to1\label{eq:-13}
\end{align}
doubly exponentially fast, as long as 
\begin{equation}
R>I_{Q}(W;X).
\end{equation}
Next we prove that there exists some $\delta,\epsilon>0$ such that
\begin{align}
 & \mathbb{P}_{\mathcal{C}_{n}}\left(D_{\infty}(\widetilde{Q}_{X^{n}}\|\widehat{Q}_{X^{n}|K_{n}\mathcal{C}_{n}}|P_{K_{n}})\leq e^{-n\delta}\right)\to1
\end{align}
doubly exponentially fast, as long as 
\begin{equation}
R>I_{Q}(W;X).
\end{equation}

For brevity, in the following we let $\mathsf{M}=e^{nR}$. According
to the definition of the R\'enyi divergence of order $\infty$, we first
have 
\begin{align}
 & e^{-D_{\infty}(\widetilde{\pi}_{X^{n}}\|\widehat{Q}_{X^{n}|K_{n}\mathcal{C}_{n}}|P_{K_{n}})}\nonumber \\
 & =\min_{x^{n}\in\mathcal{T}_{\epsilon}^{\left(n\right)},k\in[1:e^{nR_{0}}]}\frac{\widehat{Q}_{X^{n}|K_{n}\mathcal{C}_{n}}\left(x^{n}|k,\mathcal{C}_{n}\right)}{\widetilde{Q}_{X^{n}}\left(x^{n}\right)}\\
 & =\min_{x^{n}\in\mathcal{T}_{\epsilon}^{\left(n\right)},k\in[1:e^{nR_{0}}]}\widetilde{g}(x^{n}|\mathcal{C}_{n}(k)),\label{eq:-151-2}
\end{align}
where we define the function 
\begin{equation}
\widetilde{g}(x^{n}|\mathcal{C}_{n}(k)):=\sum_{m\in\calM_{n}}\frac{1}{\mathsf{M}}g(x^{n}|W^{n}(m,k))
\end{equation}
with 
\begin{equation}
\mathcal{C}_{n}(k):=\left\{ W^{n}\left(m,k\right):m\in\calM_{n}\right\} 
\end{equation}
and 
\begin{equation}
g(x^{n}|w^{n}):=\frac{1}{\widetilde{Q}_{X^{n}}\left(x^{n}\right)}P_{X^{n}|W^{n}}\left(x^{n}|w^{n}\right).
\end{equation}
Then for $w^{n}\in\mathcal{T}_{2\epsilon}^{\left(n\right)}\left(Q_{W}\right)$
and $x^{n}\in\mathcal{T}_{\epsilon}^{\left(n\right)}\left(Q_{X}\right)$,
\begin{align}
g(x^{n}|w^{n}) & =\frac{\frac{Q_{X|W}^{n}\left(x^{n}|w^{n}\right)1\left\{ x^{n}\in\mathcal{T}_{4\epsilon}^{\left(n\right)}\left(Q_{WX}|w^{n}\right)\right\} }{Q_{X|W}^{n}\left(\mathcal{T}_{4\epsilon}^{\left(n\right)}\left(Q_{WX}|w^{n}\right)|w^{n}\right)}}{\frac{Q_{X}^{n}\left(x^{n}\right)}{Q_{X}^{n}\left(\mathcal{T}_{\epsilon}^{(n)}\right)}}\\
 & \leq\frac{Q_{X}^{n}\left(\mathcal{T}_{\epsilon}^{(n)}\right)1\left\{ x^{n}\in\mathcal{T}_{4\epsilon}^{\left(n\right)}\left(Q_{WX}|w^{n}\right)\right\} }{p_{n}}\nonumber \\
 & \qquad\times e^{n\sum_{w,x}T_{w^{n}x^{n}}\left(w,x\right)\log\frac{Q\left(x|w\right)}{Q\left(x\right)}}\\
 & \leq\frac{1}{p_{n}}e^{n\left(1+4\epsilon\right)I_{Q}(W;X)}\label{eq:-4}\\
 & =:\beta_{n},\label{eq:-93-1}
\end{align}
where $p_{n}:=\min_{w^{n}\in\mathcal{T}_{2\epsilon}^{\left(n\right)}\left(Q_{W}\right)}Q_{X|W}^{n}\left(\mathcal{T}_{4\epsilon}^{\left(n\right)}\left(Q_{WX}|w^{n}\right)|w^{n}\right)$
converges to one exponentially fast as $n\to\infty$, and \eqref{eq:-4}
follows from the typical average lemma \cite{Gamal}.

Continuing \eqref{eq:-151-2}, we get for any sequence $\delta_{n}>0$,
\begin{align}
 & \mathbb{P}_{\mathcal{C}_{n}}\left(\min_{x^{n}\in\mathcal{T}_{\epsilon}^{\left(n\right)},k\in[1:e^{nR_{0}}]}\widetilde{g}(x^{n}|\mathcal{C}_{n}(k))\leq1-\delta_{n}\right)\nonumber \\
 & \leq\left|\mathcal{T}_{\epsilon}^{\left(n\right)}\right|e^{nR_{0}}\max_{\substack{x^{n}\in\mathcal{T}_{\epsilon}^{\left(n\right)},\\
k\in[1:e^{nR_{0}}]
}
}\mathbb{P}_{\mathcal{C}_{n}}\left(\widetilde{g}(x^{n}|\mathcal{C}_{n}(k))\leq1-\delta_{n}\right),\label{eq:-152-2}
\end{align}
where \eqref{eq:-152-2} follows from the union bound. Obviously,
$\left|\mathcal{T}_{\epsilon}^{\left(n\right)}\right|e^{nR_{0}}$
is only exponentially large. Therefore, if the probability vanishes
doubly exponentially fast, then $\min_{x^{n}\in\mathcal{T}_{\epsilon}^{\left(n\right)},k\in[1:e^{nR_{0}}]}\widetilde{g}(x^{n}|\mathcal{C}_{n}(k))>1-\delta_{n}$
with probability at least $1-\gamma_{n}$, where $\gamma_{n}\to0$
doubly exponentially fast as $n\to\infty$. Next we prove this.

Observe that given $x^{n}\in\mathcal{T}_{\epsilon}^{\left(n\right)}\left(Q_{X}\right),k\in[1:e^{nR_{0}}]$,
the quantities $g(x^{n}|W^{n}(m,k)),m\in\calM_{n}$ are i.i.d. random
variables with mean and variance bounded as follows. 
\begin{align}
\mu_{n} & :=\mathbb{E}_{W^{n}}\left[g(x^{n}|W^{n})\right]\\
 & =\sum_{w^{n}}\frac{Q_{W}^{n}\left(w^{n}\right)1\left\{ w^{n}\in\mathcal{T}_{2\epsilon}^{\left(n\right)}\left(Q_{W}\right)\right\} }{Q_{W}^{n}\left(\mathcal{T}_{2\epsilon}^{\left(n\right)}\left(Q_{W}\right)\right)}\nonumber \\
 & \qquad\times\frac{\frac{Q_{X|W}^{n}\left(x^{n}|w^{n}\right)1\left\{ x^{n}\in\mathcal{T}_{4\epsilon}^{\left(n\right)}\left(Q_{WX}|w^{n}\right)\right\} }{Q_{X|W}^{n}\left(\mathcal{T}_{4\epsilon}^{\left(n\right)}\left(Q_{WX}|w^{n}\right)|w^{n}\right)}}{\frac{Q_{X}^{n}\left(x^{n}\right)1\left\{ x^{n}\in\mathcal{T}_{\epsilon}^{\left(n\right)}\left(Q_{X}\right)\right\} }{Q_{X}^{n}\left(\mathcal{T}_{\epsilon}^{\left(n\right)}\left(Q_{X}\right)\right)}}\\
 & \geq Q_{X}^{n}\left(\mathcal{T}_{\epsilon}^{\left(n\right)}\left(Q_{X}\right)\right)\sum_{w^{n}}Q_{W|X}^{n}\left(w^{n}|x^{n}\right)\nonumber \\
 & \qquad\times1\left\{ w^{n}\in\mathcal{T}_{2\epsilon}^{\left(n\right)}\left(Q_{W}\right),x^{n}\in\mathcal{T}_{4\epsilon}^{\left(n\right)}\left(Q_{WX}|w^{n}\right)\right\} \label{eq:-31-1-1}\\
 & \geq Q_{X}^{n}\left(\mathcal{T}_{\epsilon}^{\left(n\right)}\left(Q_{X}\right)\right)\sum_{w^{n}}Q_{W|X}^{n}\left(w^{n}|x^{n}\right)\nonumber \\
 & \qquad\times1\left\{ \left(w^{n},x^{n}\right)\in\mathcal{T}_{2\epsilon}^{\left(n\right)}\left(Q_{WX}\right)\right\} \label{eq:-76}\\
 & \to1\textrm{ exponentially fast as }n\to\infty,\label{eq:-5}
\end{align}
where \eqref{eq:-5} follows since both $Q_{X}^{n}\left(\mathcal{T}_{\epsilon}^{\left(n\right)}\left(Q_{X}\right)\right)$
and 
\begin{equation}
q_{n}:=\min_{x^{n}\in\mathcal{T}_{\epsilon}^{\left(n\right)}}Q_{W|X}^{n}\left(\mathcal{T}_{2\epsilon}^{\left(n\right)}\left(Q_{WX}|x^{n}\right)|x^{n}\right)
\end{equation}
 converge to one (from below) exponentially fast as $n\to\infty$.
In the other direction, 
\begin{align}
\mu_{n} & \leq\sum_{w^{n}}\frac{1}{Q_{W}^{n}\left(\mathcal{T}_{2\epsilon}^{\left(n\right)}\left(Q_{W}\right)\right)p_{n}}\frac{Q_{W}^{n}\left(w^{n}\right)Q_{X|W}^{n}\left(x^{n}|w^{n}\right)}{Q_{X}^{n}\left(x^{n}\right)}\\
 & =\frac{1}{Q_{W}^{n}\left(\mathcal{T}_{2\epsilon}^{\left(n\right)}\left(Q_{W}\right)\right)p_{n}}\label{eq:-31-1-1-2}\\
 & \to1\textrm{ exponentially fast as }n\to\infty,\label{eq:-20}
\end{align}
and 
\begin{align}
\mathrm{Var}_{W^{n}}\left[g(x^{n}|W^{n})\right] & \leq\mathbb{E}_{W^{n}}\left[g(x^{n}|W^{n})^{2}\right]\leq\beta_{n}\mu_{n}.
\end{align}
We set $\delta_{n}:=e^{-n\gamma}$, where $\gamma>0$ is smaller than
the exponent of the convergence in \eqref{eq:-5}. Hence $\delta_{n}+\mu_{n}-1>0$
for sufficiently large $n$ and $\delta_{n}+\mu_{n}-1$ converges
to zero (from above) exponentially fast with the exponent $\gamma$.
Then for sufficiently large $n$, we get 
\begin{align}
 & \mathbb{P}_{\mathcal{C}_{n}}\left(\widetilde{g}(x^{n}|\mathcal{C}_{n}(k))\leq1-\delta_{n}\right)\nonumber \\
 & =\mathbb{P}_{\mathcal{C}_{n}}\Biggl(\sum_{m\in\calM_{n}}g(x^{n}|W^{n}(m,k))-\mu_{n}\mathsf{M}\nonumber \\
 & \qquad\leq\left(1-\delta_{n}-\mu_{n}\right)\mathsf{M}\Biggr)\label{eq:-94-1}\\
 & \leq\mathbb{P}_{\mathcal{C}_{n}}\Biggl(\left|\sum_{m\in\calM_{n}}g(x^{n}|W^{n}(m,k))-\mu_{n}\mathsf{M}\right|\nonumber \\
 & \qquad\geq\left(\delta_{n}+\mu_{n}-1\right)\mathsf{M}\Biggr)\\
 & \leq2\exp\left(-\frac{\frac{1}{2}\left(\delta_{n}+\mu_{n}-1\right)^{2}\mathsf{M}^{2}}{\mathsf{M}\beta_{n}\mu_{n}+\frac{1}{3}\left(\delta_{n}+\mu_{n}-1\right)\mathsf{M}\beta_{n}}\right)\label{eq:-9}\\
 & \leq2\exp\left(-\frac{3\left(\delta_{n}+\mu_{n}-1\right)^{2}\mathsf{M}}{2\left(\delta_{n}+4\mu_{n}-1\right)\beta_{n}}\right),\label{eq:-153-1}
\end{align}
where \eqref{eq:-9} follows from Bernstein's inequality \cite{boucheron2013concentration}.

Observe that $\delta_{n}+\mu_{n}-1\to0$ exponentially fast with exponent
$\gamma$, and 
\begin{equation}
\frac{\mathsf{M}}{\beta_{n}}=p_{n}e^{n\left(R-\left(1+4\epsilon\right)I_{Q}(W;X)\right)}\to\infty
\end{equation}
exponentially fast with the exponent $R-\left(1+4\epsilon\right)I_{Q}(W;X)$.
Hence \eqref{eq:-153-1} converges to zero doubly exponentially fast
as long as $R>\left(1+4\epsilon\right)I_{Q}(W;X)+2\gamma$. Since
$\epsilon,\gamma>0$ are arbitrary, such a convergence result holds
as long as $R>I_{Q}(W;X)$.

\subsection{\label{subsec:Converse-Part} Converse }

Since $X^{n}\to\left(W_{n},K_{n}\right)\to Y^{n}$ and $\left(X^{n},Y^{n}\right)\sim\pi_{XY}^{n}$,
a synthesis code forms an exact common information code \cite{yu2018on}
if we consider $\left(W_{n},K_{n}\right)$ as a common random variable.
The following converse for exact common information problem has been
proven in \cite[Appendix A-C]{yu2018on}\@.
\begin{lem}
\cite[Appendix A-C]{yu2018on} For a sequence of  random triples $\left\{ \left(X^{n},Y^{n},Z_{n}\right)\right\} $
such that $\left(X^{n},Y^{n}\right)\sim\pi_{XY}^{n}$ and $X^{n}\to Z_{n}\to Y^{n}$,
we have 
\begin{align}
\frac{1}{n}H(Z_{n}) & \ge-\frac{1}{n}H(X^{n}Y^{n}|Z_{n})+\frac{1}{n}\sum_{z}P_{Z_{n}}(z)\nonumber \\
 & \quad\times\mathcal{H}(P_{X^{n}|Z_{n}=z},P_{Y^{n}|Z_{n}=z}\|\pi_{XY}^{n})+o(1)
\end{align}
where $o(1)$ denotes a term that vanishes as $n\to\infty$.
\end{lem}
By the lemma above, we have
\begin{align}
R_{0}+R & \ge-\frac{1}{n}H(X^{n}Y^{n}|W)+\frac{1}{n}\sum_{w}P(w)\nonumber \\
 & \quad\times\mathcal{H}(P_{X^{n}|W=w},P_{Y^{n}|W=w}\|\pi_{XY}^{n}),
\end{align}
where $W:=\left(W_{n},K_{n}\right)$.

On the other hand, 
\begin{align}
R & \ge\frac{1}{n}H(W_{n}|K_{n})\\
 & \ge\frac{1}{n}I(X^{n};W_{n}|K_{n})\\
 & =\frac{1}{n}I(X^{n};W_{n}K_{n})\label{eq:-37}
\end{align}
where \eqref{eq:-37} follows since $X^{n}$ is independent of $K_{n}$.

\section{\label{sec:singleletter}Proof of Theorem \ref{thm:singleletter}}

The inner bound has been proved in Appendix \ref{subsec:achievability}.
On the other hand, $\mathcal{R}_{\mathrm{Exact}}(\pi_{XY})\subseteq\mathcal{R}_{\mathrm{Cuff}}(\pi_{XY})$
follows from the following argument. By the definition of the maximal
cross-entropy of couplings, we have 
\begin{equation}
\sum_{w}P(w)\mathcal{H}(P_{X^{n}|W=w},P_{Y^{n}|W=w}\|\pi_{XY}^{n})\ge nH_{\pi}(XY),
\end{equation}
we have $\mathcal{R}(\pi_{XY}^{n})\subseteq\mathcal{R}_{\mathrm{Cuff}}(\pi_{XY}^{n})$.
Cuff \cite{Cuff} showed that $\frac{1}{n}\mathcal{R}_{\mathrm{Cuff}}(\pi_{XY}^{n})=\mathcal{R}_{\mathrm{Cuff}}(\pi_{XY})$.
Hence by using the multi-letter characterization in Theorem \ref{thm:multiletter},
we have 
\begin{equation}
\mathcal{R}_{\mathrm{Exact}}(\pi_{XY})\subseteq\mathcal{R}_{\mathrm{Cuff}}(\pi_{XY}).
\end{equation}
Hence we only need to prove the outer bound $\mathcal{R}^{(\mathrm{o})}(\pi_{XY})$.

Denote $J\sim P_{J}:=\mathrm{Unif}[1:n]$ as a time index independent
of $(W_{n},K_{n},X^{n},Y^{n})$. Denote $W:=W_{n}K_{n}JX^{J-1}Y^{J-1},X:=X_{J},Y:=Y_{J}$.
Since $X^{n}\to W_{n}K_{n}\to Y^{n}$ and $\left(X^{n},Y^{n}\right)\sim\pi_{XY}^{n}$,
a channel synthesis code is also an exact common information code
\cite{yu2018on}. The multi-letter expression of the sum-rate in $\mathcal{R}(\pi_{XY}^{n})$
is lower bounded by the following single-letter expression.
\begin{lem}
\cite[Theorem 2]{yu2018on} For a sequence of  random triples $\left\{ \left(X^{n},Y^{n},Z_{n}\right)\right\} $
such that $\left(X^{n},Y^{n}\right)\sim\pi_{XY}^{n}$ and $X^{n}\to Z_{n}\to Y^{n}$,
we have 
\begin{align}
 & -\frac{1}{n}H(X^{n}Y^{n}|Z_{n})+\frac{1}{n}\sum_{z}P_{Z_{n}}(z)\nonumber \\
 & \quad\times\mathcal{H}(P_{X^{n}|Z_{n}=z},P_{Y^{n}|Z_{n}=z}\|\pi_{XY}^{n})\nonumber \\
 & \geq-H(XY|W)+\min_{Q_{WW'}\in C(P_{W},P_{W})}\sum_{w,w'}Q_{WW'}(w,w')\nonumber \\
 & \quad\times\mathcal{H}(P_{X|W=w},P_{Y|W=w'}\|\pi_{XY}).
\end{align}
where $W:=Z_{n}JX^{J-1}Y^{J-1},X:=X_{J}$, $Y:=Y_{J}$, and $J\sim P_{J}:=\mathrm{Unif}[1:n]$
denotes a random time index independent of $(Z_{n},X^{n},Y^{n})$.
\end{lem}
By the lemma above, the multi-letter expression of the sum rate in
$\mathcal{R}(\pi_{XY}^{n})$ can be lower bounded as 
\begin{align}
R_{0}+R & \ge-H(XY|W)\nonumber \\
 & \quad+\min_{Q_{WW'}\in C(P_{W},P_{W})}\sum_{w,w'}Q_{WW'}(w,w')\nonumber \\
 & \quad\times\mathcal{H}(P_{X|W=w},P_{Y|W=w'}\|\pi_{XY}).
\end{align}

On the other hand, observe that 
\begin{align}
 & P_{W_{n}K_{n}X^{i}Y^{i-1}}\nonumber \\
 & =P_{W_{n}K_{n}}P_{X^{i}|W_{n}K_{n}}P_{Y^{i-1}|W_{n}K_{n}}\\
 & =P_{W_{n}K_{n}}P_{X^{i-1}|W_{n}K_{n}}P_{X_{i}|W_{n}K_{n}X^{i-1}}P_{Y^{i-1}|W_{n}K_{n}}.
\end{align}
Hence $X_{i}\rightarrow W_{n}K_{n}X^{i-1}\rightarrow Y^{i-1}$ forms
a Markov chain. We then get 
\begin{align}
R & \ge\frac{1}{n}H(W_{n}|K_{n})\\
 & \ge\frac{1}{n}I(X^{n};W_{n}|K_{n})\\
 & =\frac{1}{n}I(X^{n};W_{n}K_{n})\\
 & =\frac{1}{n}\sum_{i=1}^{n}I(X_{i};W_{n}K_{n}|X^{i-1})\\
 & =\frac{1}{n}\sum_{i=1}^{n}I(X_{i};X^{i-1}W_{n}K_{n})\\
 & =\frac{1}{n}\sum_{i=1}^{n}I(X_{i};X^{i-1}Y^{i-1}W_{n}K_{n})\label{eq:-53}\\
 & =I(X_{J};X^{J-1}Y^{J-1}W_{n}K_{n}|J)\\
 & =I(X_{J};X^{J-1}Y^{J-1}W_{n}K_{n}J)\\
 & =I(X;W),
\end{align}
where \eqref{eq:-53} follows since $X_{i}\rightarrow W_{n}K_{n}X^{i-1}\rightarrow Y^{i-1}$
forms a Markov chain.

\section{\label{sec:DSBS}Proof of Theorem \ref{thm:DSBS}}

Inner Bound: Set $X=W\oplus A$ and $Y=W\oplus B$ with $W\sim\mathrm{Bern}(\frac{1}{2})$,
$A\sim\mathrm{Bern}(a)$, and $B\sim\mathrm{Bern}(b)$ mutually independent,
where $b:=\frac{p-a}{1-2a},a\in(0,p)$. That is, $a\overline{b}+\overline{a}b=p$.
Since by the assumption $p<\frac{1}{2}$, we have $a,b<\frac{1}{2}$.
From Example \ref{exa:-Consider-a}, we have that 
\begin{align}
 & \mathcal{H}(P_{X|W=w},P_{Y|W=w}\|\pi_{XY})\nonumber \\
 & =\log\frac{1}{\alpha_{0}}+\left(a+b\right)\log\frac{\alpha_{0}}{\beta_{0}}.
\end{align}
Hence we have 
\begin{align}
 & \mathcal{R}(\pi_{XY})\nonumber \\
 & \subseteq\left\{ \begin{array}{rcl}
\left(R,R_{0}\right)\in\mathbb{R}_{\ge0}^{2} & : & a\in(0,p),b:=\frac{p-a}{1-2a},\\
R & \ge & 1-H_{2}(a),\\
R_{0}+R & \ge & -H_{2}(a)-H_{2}(b)\\
 &  & +\log\frac{1}{\alpha_{0}}+\left(a+b\right)\log\frac{\alpha_{0}}{\beta_{0}}
\end{array}\right\} .
\end{align}

Outer Bound: We adopt similar techniques as the ones used by Wyner
\cite{Wyner}. Denote 
\begin{align}
\alpha(w) & :=\mathbb{P}\left(X=0|W=w\right)\\
\beta(w) & :=\mathbb{P}\left(Y=0|W=w\right).
\end{align}
Hence $P_{XY}=\pi_{XY}$ implies 
\begin{align}
\mathbb{E}\alpha(W) & =\mathbb{P}\left(X=0\right)=\frac{1}{2}\\
\mathbb{E}\beta(W) & =\mathbb{P}\left(Y=0\right)=\frac{1}{2}\\
\mathbb{E}\alpha(W)\beta(W) & =\mathbb{P}\left(X=0,Y=0\right)=\alpha_{0}.
\end{align}
From Example \ref{exa:-Consider-a}, we have that 
\begin{align}
 & \mathcal{H}(P_{X|W=w},P_{Y|W=w'}\|\pi_{XY})\nonumber \\
 & =\log\frac{1}{\alpha_{0}}+\min\{\alpha(w)+\beta(w'),\overline{\alpha(w)}+\overline{\beta(w')}\}\log\frac{\alpha_{0}}{\beta_{0}}\\
 & \geq\log\frac{1}{\alpha_{0}}+\left(\min\{\alpha(w),\overline{\alpha(w)}\}+\min\{\beta(w'),\overline{\beta(w')}\right)\nonumber \\
 & \qquad\qquad\times\log\frac{\alpha_{0}}{\beta_{0}}.
\end{align}

Define $\alpha'(W):=\left|\alpha(W)-\frac{1}{2}\right|,\beta'(W):=\left|\beta(W)-\frac{1}{2}\right|$,
$\delta(W):=\alpha'^{2}(W)$, $\gamma(W):=\beta'^{2}(W)$, $a:=\frac{1}{2}-\sqrt{\mathbb{E}\delta(W)}$,
and $b:=\frac{1}{2}-\sqrt{\mathbb{E}\gamma(W)}$. Then we have that
\begin{align}
 & R_{0}+R\nonumber \\
 & \geq-\mathbb{E}H_{2}(\alpha(W))-\mathbb{E}H_{2}(\beta(W))+\log\frac{1}{\alpha_{0}}\nonumber \\
 & \qquad+\left(\mathbb{E}\min\{\alpha(W),\overline{\alpha(W)}\}+\mathbb{E}\min\{\beta(W),\overline{\beta(W)}\}\right)\nonumber \\
 & \qquad\qquad\times\log\frac{\alpha_{0}}{\beta_{0}}\\
 & =-\mathbb{E}H_{2}\left(\frac{1}{2}+\alpha'(W)\right)-\mathbb{E}H_{2}\left(\frac{1}{2}+\beta'(W)\right)+\log\frac{1}{\alpha_{0}}\nonumber \\
 & \qquad+\left(\mathbb{E}\left(\frac{1}{2}-\alpha'(W)\right)+\mathbb{E}\left(\frac{1}{2}-\beta'(W)\right)\right)\log\frac{\alpha_{0}}{\beta_{0}}\label{eq:-8}\\
 & \geq-H_{2}\left(\frac{1}{2}+\sqrt{\mathbb{E}\delta(W)}\right)-H_{2}\left(\frac{1}{2}+\sqrt{\mathbb{E}\gamma(W)}\right)+\log\frac{1}{\alpha_{0}}\nonumber \\
 & \qquad+\left(\frac{1}{2}-\sqrt{\mathbb{E}\delta(W)}+\frac{1}{2}-\sqrt{\mathbb{E}\gamma(W)}\right)\log\frac{\alpha_{0}}{\beta_{0}}\label{eq:-54}\\
 & =-H_{2}\left(a\right)-H_{2}\left(b\right)+\log\frac{1}{\alpha_{0}}+\left(a+b\right)\log\frac{\alpha_{0}}{\beta_{0}},\label{eq:-60}
\end{align}
where \eqref{eq:-54} follows from the fact both $x\mapsto H_{2}\left(\frac{1}{2}+\sqrt{x}\right)$
for $x\in[0,\frac{1}{4}]$ and $x\mapsto\sqrt{x}$ for $x\ge0$ are
concave functions \cite[Prop. 3.3]{Wyner}. Similarly, 
\begin{align}
R & \ge I(W;X)\\
 & =1-\mathbb{E}H_{2}(\alpha(W))\\
 & =1-\mathbb{E}H_{2}\left(\frac{1}{2}+\alpha'(W)\right)\\
 & \geq1-H_{2}\left(\frac{1}{2}+\sqrt{\mathbb{E}\delta(W)}\right)\\
 & =1-H_{2}\left(a\right).\label{eq:-66}
\end{align}
On the other hand, 
\begin{align}
 & \begin{cases}
\mathbb{E}\alpha(W)=\frac{1}{2}\\
\mathbb{E}\beta(W)=\frac{1}{2}\\
\mathbb{E}\alpha(W)\beta(W)=\alpha_{0}
\end{cases}\nonumber \\
\Rightarrow & \begin{cases}
0\leq\alpha'(w),\beta'(w)\leq\frac{1}{2}\\
\mathbb{E}\alpha'(W)\beta'(W)\geq\alpha_{0}-\frac{1}{4}
\end{cases}\\
\Rightarrow & \begin{cases}
0\leq\delta(W),\gamma(W)\leq\frac{1}{4}\\
\mathbb{E}\sqrt{\delta(W)\gamma(W)}\geq\alpha_{0}-\frac{1}{4}
\end{cases}\\
\Rightarrow & a\overline{b}+\overline{a}b\leq p,\label{eq:-56}
\end{align}
where \eqref{eq:-56} follows since by the Cauchy--Schwarz inequality,
we have 
\begin{align}
 & a\overline{b}+\overline{a}b\nonumber \\
 & =a+b-2ab\\
 & =1-\sqrt{\mathbb{E}\delta(W)}-\sqrt{\mathbb{E}\gamma(W)}\nonumber \\
 & \quad-\left(\frac{1}{2}-\sqrt{\mathbb{E}\delta(W)}-\sqrt{\mathbb{E}\gamma(W)}+2\sqrt{\mathbb{E}\delta(W)}\sqrt{\mathbb{E}\gamma(W)}\right)\\
 & =\frac{1}{2}-2\sqrt{\mathbb{E}\delta(W)\mathbb{E}\gamma(W)}\\
 & \leq\frac{1}{2}-2\mathbb{E}\sqrt{\delta(W)\gamma(W)}\label{eq:-57}\\
 & \leq p.
\end{align}
Combining \eqref{eq:-60}, \eqref{eq:-66}, and \eqref{eq:-56} yields
the desired result.

\subsection*{Acknowledgements}
The authors would like to thank the Associate Editor Prof.\  Amin Gohari and the three reviewers for their extensive, constructive and helpful feedback to improve the manuscript.

\bibliographystyle{unsrt}
\bibliography{ref}

\begin{IEEEbiographynophoto}{Lei Yu} received the B.E. and Ph.D. degrees, both in electronic engineering, from University of Science and Technology of China (USTC) in 2010 and 2015, respectively. From 2015 to 2017, he was a postdoctoral researcher at the Department of Electronic Engineering and Information Science (EEIS), USTC. Currently, he is a research fellow at the Department of Electrical and Computer Engineering, National University of Singapore. His research interests lie in the intersection of information theory, probability theory, and combinatorics.  \end{IEEEbiographynophoto}
\begin{IEEEbiographynophoto}{Vincent Y.\ F.\ Tan} (S'07-M'11-SM'15) was born in Singapore in 1981. He is currently a Dean's Chair Associate Professor in the Department of Electrical and Computer Engineering and the Department of Mathematics at the National University of Singapore (NUS). He received the B.A.\ and M.Eng.\ degrees in Electrical and Information Sciences from Cambridge University in 2005 and the Ph.D.\ degree in Electrical Engineering and Computer Science (EECS) from the Massachusetts Institute of Technology (MIT) in 2011. His research interests include information theory, machine learning, and statistical signal processing.
Dr.\ Tan received the MIT EECS Jin-Au Kong outstanding doctoral thesis prize in 2011, the NUS Young Investigator Award in 2014, the NUS Engineering Young Researcher Award in 2018, and the Singapore National Research Foundation (NRF) Fellowship (Class of 2018). He is also an IEEE Information Theory Society Distinguished Lecturer for 2018/9. He has authored a research monograph on {\em ``Asymptotic Estimates in Information Theory with Non-Vanishing Error Probabilities''} in the Foundations and Trends in Communications and Information Theory Series (NOW Publishers). He is currently serving as an Associate Editor of the IEEE Transactions on Signal Processing.  \end{IEEEbiographynophoto} 
\end{document}

%% file: preamble.tex
\usepackage[mathscr]{eucal}
\usepackage{fixltx2e}
\usepackage{array}
\usepackage{verbatim}
\usepackage{bm}
\usepackage{algorithmic}
\usepackage{algorithm}
\usepackage{verbatim}
\usepackage{textcomp}
\usepackage{mathrsfs}
\usepackage{epstopdf}

\newcommand{\openone}{\leavevmode\hbox{\small1\normalsize\kern-.33em1}}

\catcode`~=11 \def\UrlSpecials{\do\~{\kern -.15em\lower .7ex\hbox{~}\kern .04em}} \catcode`~=13

\allowdisplaybreaks[4]

\newcommand{\calK}{\mathcal{K}}

\newcommand{\calM}{\mathcal{M}}

\newcommand{\calP}{\mathcal{P}}

\newcommand{\calX}{\mathcal{X}}

\DeclareMathAlphabet{\mathbsf}{OT1}{cmss}{bx}{n}
\DeclareMathAlphabet{\mathssf}{OT1}{cmss}{m}{sl}

\DeclareSymbolFont{bsfletters}{OT1}{cmss}{bx}{n}
\DeclareSymbolFont{ssfletters}{OT1}{cmss}{m}{n}
\DeclareMathSymbol{\bsfGamma}{0}{bsfletters}{'000}
\DeclareMathSymbol{\ssfGamma}{0}{ssfletters}{'000}
\DeclareMathSymbol{\bsfDelta}{0}{bsfletters}{'001}
\DeclareMathSymbol{\ssfDelta}{0}{ssfletters}{'001}
\DeclareMathSymbol{\bsfTheta}{0}{bsfletters}{'002}
\DeclareMathSymbol{\ssfTheta}{0}{ssfletters}{'002}
\DeclareMathSymbol{\bsfLambda}{0}{bsfletters}{'003}
\DeclareMathSymbol{\ssfLambda}{0}{ssfletters}{'003}
\DeclareMathSymbol{\bsfXi}{0}{bsfletters}{'004}
\DeclareMathSymbol{\ssfXi}{0}{ssfletters}{'004}
\DeclareMathSymbol{\bsfPi}{0}{bsfletters}{'005}
\DeclareMathSymbol{\ssfPi}{0}{ssfletters}{'005}
\DeclareMathSymbol{\bsfSigma}{0}{bsfletters}{'006}
\DeclareMathSymbol{\ssfSigma}{0}{ssfletters}{'006}
\DeclareMathSymbol{\bsfUpsilon}{0}{bsfletters}{'007}
\DeclareMathSymbol{\ssfUpsilon}{0}{ssfletters}{'007}
\DeclareMathSymbol{\bsfPhi}{0}{bsfletters}{'010}
\DeclareMathSymbol{\ssfPhi}{0}{ssfletters}{'010}
\DeclareMathSymbol{\bsfPsi}{0}{bsfletters}{'011}
\DeclareMathSymbol{\ssfPsi}{0}{ssfletters}{'011}
\DeclareMathSymbol{\bsfOmega}{0}{bsfletters}{'012}
\DeclareMathSymbol{\ssfOmega}{0}{ssfletters}{'012}

\DeclareMathOperator{\supp}{supp}

